%% file: usenix.tex
\documentclass[letterpaper,twocolumn,10pt]{article}
\usepackage{usenix,epsfig,endnotes}
\usepackage{color}
\usepackage{overpic}
\usepackage{amsmath}
\usepackage{amssymb}
\usepackage{pifont}
\usepackage{multirow}
\usepackage{enumitem}
\usepackage{subfigure}
\usepackage{makecell}
\usepackage{caption}
\usepackage{xspace}
\usepackage{etoolbox}

\usepackage[available,functional,reproduced]{usenixbadges}

\patchcmd{\thebibliography}{\chapter*}{\section*}{}{}

\newcommand{\nomad}{\textsc{Nomad}\xspace} 
\begin{document}

\date{}

\title{\Large \nomad: Non-Exclusive Memory Tiering via Transactional Page Migration}
\author{
 {\rm Lingfeng Xiang, Zhen Lin, Weishu Deng, Hui Lu, Jia Rao, Yifan Yuan$^\dagger$, Ren Wang$^\dagger$}\\
 The University of Texas at Arlington, $^\dagger$Intel Labs
 }

\maketitle

\thispagestyle{empty}

\subsection*{Abstract}
With the advent of byte-addressable memory devices, such as CXL memory, persistent memory, and storage-class memory, tiered memory systems have become a reality. Page migration is the {\em de facto} method within operating systems for managing tiered memory. It aims to bring hot data whenever possible into fast memory to optimize the performance of data accesses while using slow memory to accommodate data spilled from fast memory. While the existing research has demonstrated the effectiveness of various optimizations on page migration, it falls short of addressing a fundamental question: Is exclusive memory tiering, in which a page is either present in fast memory or slow memory, but not both simultaneously, the optimal strategy for tiered memory management?

We demonstrate that page migration-based exclusive memory tiering suffers significant performance degradation when fast memory is under pressure. In this paper, we propose {\em non-exclusive} memory tiering, a page management strategy that retains a copy of pages recently promoted from slow memory to fast memory to mitigate memory thrashing. To enable non-exclusive memory tiering, we develop \textsc{Nomad}, a new page management mechanism for Linux that features {\em transactional page migration} and {\em page shadowing}. \textsc{Nomad} helps remove page migration off the critical path of program execution and makes migration completely asynchronous. Evaluations with carefully crafted micro-benchmarks and real-world applications show that \textsc{Nomad} is able to achieve up to 6x performance improvement over the state-of-the-art transparent page placement (TPP) approach in Linux when under memory pressure. We also compare \textsc{Nomad} with a recently proposed hardware-assisted, access sampling-based page migration approach and demonstrate \textsc{Nomad}'s strengths and potential weaknesses in various scenarios. 

\input{introduction}
\input{design}

\input{evaluation}

{\footnotesize \bibliographystyle{acm}
\bibliography{reference}
}
\include{osdi24_ae_appendix_template}

\end{document}

%% file: introduction.tex
\section{Introduction}
As new memory devices, such as high bandwidth memory (HBM)~\cite{hbm, jun2017hbm}, DRAM, persistent memory~\cite{optane, p2cache}, Compute Express Link (CXL)-based memory~\cite{cxl, Maruf_23asplos_tpp, memtis}, and storage-class memory~\cite{optane_study, xiang2022characterizing} continue to emerge, future computer systems are anticipated to feature multiple tiers of memory with distinct characteristics, such as speed, size, power, and cost. Tiered memory management aims to leverage the strength of each memory tier to optimize the overall data access latency and bandwidth. Central to tiered memory management is {\em page management} within operating systems (OS), including page allocation, placement, and migration. Efficient page management in the OS is crucial for optimizing memory utilization and performance while maintaining transparency for user applications. 

Traditionally, the memory hierarchy consists of storage media with at least one order of magnitude difference in performance. For example, in the two-level memory hierarchy assumed by commercial operating systems for decades, DRAM and disks differ in latency, bandwidth, and capacity by 2-3 orders of magnitude. Therefore, the sole goal of page management is to keep hot pages in, and maximize the hit rate of the ``performance'' tier (DRAM), and migrate (evict) cold pages to the ``capacity'' tier (disk) when needed. As new memory devices emerge, the performance gap in the memory hierarchy narrows. Evaluations on Intel's Optane persistent memory ~\cite{pmperf} and CXL memory~\cite{Sun2023DemystifyingCM} reveal that these new memory technologies can achieve comparable performance to DRAM in both latency and bandwidth, within a range of 2-3x. As a result, the assumption of the performance gap, which has guided the design of OS page management for decades, may not hold. It is no longer beneficial to promote a hot page to the performance tier if the migration cost is too high. 

Furthermore, unlike disks which must be accessed through the file system as a block device, new memory devices are byte-addressable and can be directly accessed by the processor via ordinary {\tt load} and {\tt store} instructions. Therefore, for a warm page on the capacity tier, accessing the page directly and avoiding migration to the performance tier could be a better option. Most importantly, while the performance of tiered memory remains hierarchical, the hardware is no longer hierarchical. Both the Optane persistent memory and CXL memory appear to the processor as a CPUless memory node and thus can be used by the OS as ordinary DRAM.

These unique challenges facing emerging tiered memory systems have inspired research on improving page management in the OS. Much focus has been on expediting page migrations between memory tiers. Nimble~\cite{Yan_19asplos_nimble} improves page migration by utilizing transparent huge pages (THP), multi-threaded migration of a page, and concurrent migration of multiple pages. Transparent page placement (TPP)~\cite{Maruf_23asplos_tpp} extends the existing NUMA balancing scheme in Linux to support asynchronous page demotion and synchronous page promotion between fast and slow memory. Memtis~\cite{memtis} and TMTS~\cite{Padmapriya_23asplos} use hardware performance counters to mitigate the overhead of page access tracking and use background threads to periodically and asynchronously promote pages.

However, these approaches have two fundamental limitations. {\em First}, the existing page management for tiered memory assumes that memory tiers are {\em exclusive} to each other -- hot pages are allocated or migrated to the performance tier while cold pages are demoted to the capacity tier. Therefore, each page is only present in one tier. As memory tiering seeks to explore the tradeoff between performance and capacity, the working set size of workloads that benefit most from tiered memory systems likely exceeds the capacity of the performance tier. Exclusive memory tiering inevitably leads to excessive hot-cold page swapping or memory thrashing when the performance tier is not large enough to hold hot data. 

{\em Second}, there is a lack of an efficient page migration mechanism to support tiered memory management. As future memory tiers are expected to be addressable by the CPU, page migrations are similar to serving minor page faults and involve three steps: 1) unmap a page from the page table; 2) copy the page content to a different tier; 3) remap the page on the page table, pointing to the new memory address. Regardless of whether page migration is done synchronously upon accessing a hot page in the slower capacity tier or asynchronously in the background, the 3-step migration process is expensive. During migration, an unmapped page cannot be accessed by user programs. If page migration is done frequently, e.g., due to memory thrashing, user-perceived bandwidth, including accesses to the migrating pages, is significantly lower (up to 95\% lower) than the peak memory bandwidth~\cite{Yan_19asplos_nimble}. 

This paper advocates {\em non-exclusive} memory tiering that allows a subset of pages on the performance tier to have shadow copies on the capacity tier~\footnote{We assume that page migrations only occur between two adjacent tiers if there are more than two memory tiers.}. Note that non-exclusive tiering is different from {\em inclusive} tiering which strictly uses the performance tier as a cache of the capacity tier. The most important benefit is that under memory pressure, page demotion is made less expensive by simply remapping a page if it is not dirty and its shadow copy exists on the capacity tier. This allows for smooth performance transition when memory demand exceeds the capacity of the performance tier. 

To reduce the cost of page migration, especially for promotion, this paper proposes {\em transactional page migration} (TPM), a novel mechanism to enable page access during migration. Unlike current page migrations, TPM starts page content copy without unmapping the page from the capacity tier so that the migrating page is still accessible by user programs. After page content is copied to a new page on the performance tier, TPM checks whether the page has been dirtied during the migration. If so, the page migration (i.e., the transaction) is invalidated and the copied page is discarded. Failed page migrations will be retried at a later time. If successful, the copied new page is mapped in the page table and the old page is unmapped, becoming a shadow copy of the new page.  

We have developed \textsc{Nomad}, a new page management framework for tired memory that integrates non-exclusive memory tiering and transactional page migration. \textsc{Nomad} safeguards page allocation to prevent out-of-memory (OOM) errors due to page shadowing. When the capacity tier is under memory pressure, \textsc{Nomad} prioritizes the reclamation of shadow pages before evicting ordinary pages. We have implemented a prototype of \textsc{Nomad} in Linux and performed a thorough evaluation on four different platforms, including an FPGA-based CXL prototype, a persistent memory system, and a pre-market, commercial CXL system. Experimental results show that, compared to two representative page management schemes: TPP and Memtis, \textsc{Nomad} achieves up to 6x performance improvement over TPP during memory thrashing and consistently outperforms Memtis by as much as 130\% when the working set size fits into fast memory.

\section{Motivation and Related Work}
We introduce the background of page management in tiered memory systems and use TPP~\cite{Maruf_23asplos_tpp}, a state-of-the-art page placement system designed for CXL-enabled tiered memory, as a motivating example to highlight the main limitations of current page management approaches.

\begin{figure}[t]
	\centerline{\includegraphics[width=0.40\textwidth]{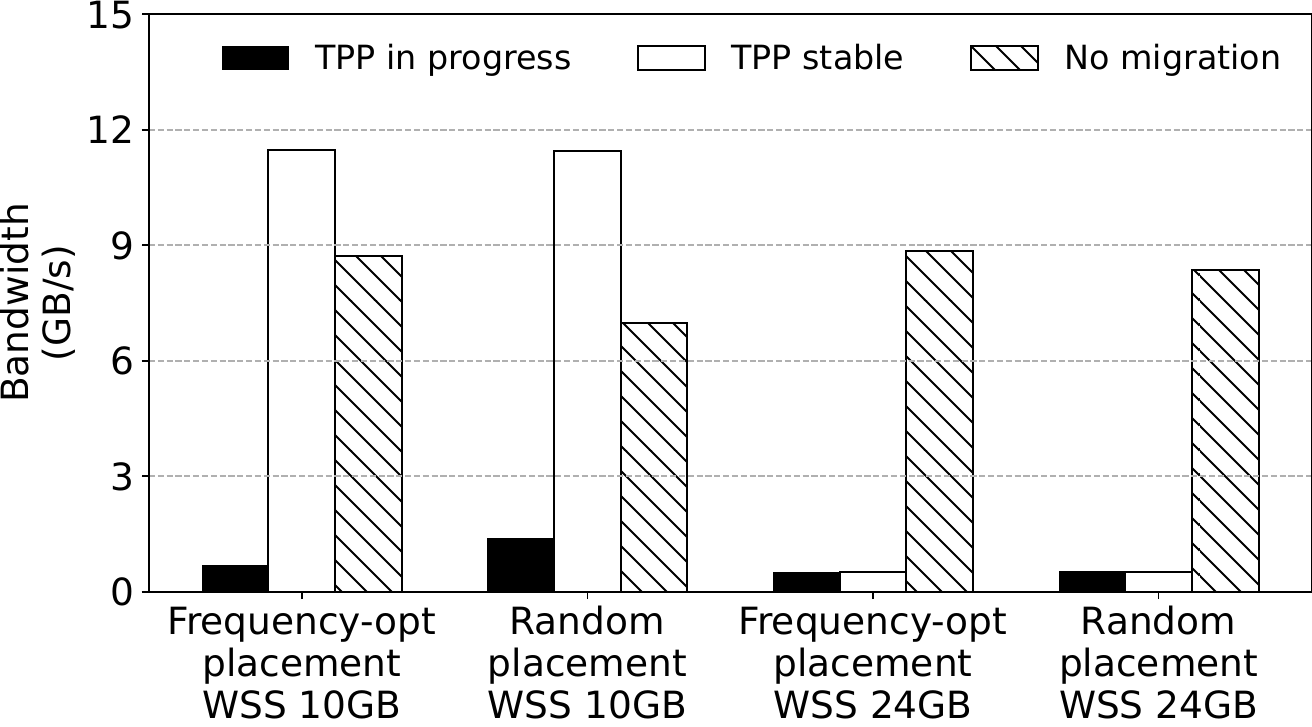}}
	\caption{\small The comparison of achieved memory bandwidth in a micro-benchmark due to different phases in TPP and a baseline approach that disables page migration. Higher is better performance.}
	\label{fig:TppVsNoMigrationPerformance}
\end{figure}

\subsection{Memory Tiering}
\noindent{\bf Caching and tiering} are two traditional software approaches to manage a memory, or storage hierarchy, consisting of various types of storage media (e.g., CPU caches, DRAM, and hard disks) differing in performance, capacity, and cost. Without loss of generality, we consider a two-level memory hierarchy with 1) a {\em performance} tier (i.e., the fast tier), backed with smaller, faster, but more expensive storage media; and 2) a {\em capacity} tier (i.e., the slow tier) with larger, slower, and cheaper storage media. For {\em caching}, data is stored in the capacity tier, and copies of frequently accessed or ``hot'' data are strategically replicated to the performance tier. For {\em tiering}, new data is first allocated to the performance tier and remains there if it is frequently accessed, while less accessed data may be relegated to the capacity tier when needed. At any moment, data resides exclusively in one of the tiers but not both. Essentially, caching operates in an {\em inclusive} page placement mode and retains pages in their original locations, only temporarily storing a copy in the performance tier for fast access. Conversely, tiering operates in an {\em exclusive mode}, actively relocating pages across various memory/storage mediums.

\smallskip

\noindent{\bf Diverse memory/storage devices}, such as high bandwidth memory (HBM)~\cite{hbm}, CXL-based memory~\cite{cxl}, persistent memory (PM)~\cite{optane}, 
and fast, byte-addressable NVMe SSDs~\cite{pcieovercxl}, have emerged recently. While they still make a tradeoff between speed, size, and cost, the gap between their performance narrows. 
For example, Intel Optane DC persistent memory (PM), available in a DIMM package on the memory bus enabling programs to directly access data from the CPU using \texttt{load} and \texttt{store} instructions, provides (almost) an order of magnitude higher capacity than DRAM (e.g., 8x) and offers performance within a range of 2-3x of DRAM, e.g., write latency as low as 80 {\em ns} and read latency around 170 {\em ns}~\cite{pmperf}. More recently, compute express link (CXL), an open-standard interconnect technology based on PCI Express (PCIe)~\cite{cxl}, provides a memory-like, byte-addressable interface (i.e., via the \texttt{CXL.mem} protocol) for connecting diverse memory devices (e.g., DRAM, PM, GPUs, and smartNICs). 
Real-world CXL memory offers comparable memory access latency (<2x) and throughput ($\sim$50\%) to ordinary DRAM~\cite{Sun2023DemystifyingCM}. 

From the perspective of OS memory management, CXL memory or PM appears to be a remote, CPUless memory node, similar to a multi-socket non-uniform memory access (NUMA) node. State-of-the-art tiered memory systems, such as TPP~\cite{Maruf_23asplos_tpp}, Memtis~\cite{memtis}, Nimble~\cite{Yan_19asplos_nimble}, and  AutoTiering\cite{Jonghyeon_21fast}, all adopt {\em tiering} to {\em exclusively} manage data on different memory tiers. Unlike the traditional two-level memory hierarchy involving DRAM and disks, in which DRAM acts as a cache for the much larger storage tier, current CXL memory tiering treats CXL memory as an extension of local DRAM. While exclusive memory tiering avoids data redundancy, it necessitates data movement between memory tiers to optimize the performance of data access, i.e., promoting hot data to the fast tier and demoting cold data to the slow tier. Given that all memory tiers are byte-addressable by the CPU and the performance gap between tiers narrows, it remains to be seen whether exclusive tiering is the optimal strategy considering the cost of data movement. 

We evaluate the performance of transparent page placement (TPP)~\cite{Maruf_23asplos_tpp}, a state-of-the-art and the default tiered memory management in Linux. Figure~\ref{fig:TppVsNoMigrationPerformance} shows the bandwidth of a micro-benchmark that accesses a configurable working set size (WSS) following a Zipfian distribution in a CXL-based tiered memory system. 
More details of the benchmark and the hardware configurations can be found in Section~\ref{sec:eval}. 
We compare the performance of TPP while it actively migrates pages between tiers for promotion and demotion (denoted as {\em TPP in progress}) and when it has finished page relocation ({\em TPP stable}) with that of a baseline that disables page migration ({\em no migration}). 
The baseline does not optimize page placement and directly accesses hot pages from the slow tier. The tiered memory testbed is configured with 16GB fast memory (local DRAM) and 16GB slow memory (remote CXL memory). We vary the WSS to fit in (e.g., 10GB) and exceed (e.g., 24GB) fast memory capacity. Note that the latter requires continuous page migrations between tiers since hot data spills into slow memory. Additionally, we explore two initial data placement strategies in the benchmark. First, the benchmark pre-allocates 10GB of data in fast memory to emulate the existing memory usage from other applications. {\em Frequency-opt} is an allocation strategy that places pages according to the descending order of their access frequencies (hotness). Thus, the hottest pages are initially placed in fast memory until the WSS spills into slow memory. In contrast, {\em Random} employs a random allocation policy and may place cold pages initially in fast memory.

We have important observations from results in Figure~\ref{fig:TppVsNoMigrationPerformance}. First, page migration in TPP incurs significant degradation in application performance. When WSS fits in fast memory, {\em TPP stable}, which has successfully migrated all hot pages to fast memory, achieves more than an order of magnitude higher bandwidth than {\em TPP in progress}. Most importantly, {\em no migration} is consistently and substantially better than {\em TPP in progress}, suggesting that the overhead of page migration outweighs its benefit until the migration is completed. Second, TPP never reaches a stable state and enters memory thrashing when WSS is larger than the capacity of fast memory. Third, page migration is crucial to achieving optimal performance if it is possible to move all hot data to fast memory and the initial placement is sub-optimal, as evidenced by the wide gap between {\em TPP stable} and {\em no migration} in the 10GB WSS and random placement test.

\subsection{Page Management} 
\label{sec:page-management}
In this section, we delve into the design of page management in Linux and analyze its overhead during page migration. We focus our discussions on 1) how to effectively track memory accesses and identify hot pages, and 2) the mechanism to migrate a page between memory tiers.


\smallskip
\noindent{\bf Tracking memory access} can be conducted by software (via the kernel) and/or with hardware assistance. Specifically, the kernel can keep track of page accesses via page faults~\cite{autoNUMA, Maruf_23asplos_tpp, Jonghyeon_21fast}, scanning page tables~\cite{Bergman_22ismm, autoNUMA, Adnan_22hpca, Yan_19asplos_nimble, thermostat}, or both.
Capturing each memory access for precise tracking can be expensive. Page fault-based tracking traps memory accesses to selected pages (i.e., whose page table entry permissions are set to {\tt no access}) via hint (minor) page faults. Thus, it allows the kernel to accurately measure the {\em recency} and {\em frequency} of these pages. However, invoking a page fault on every memory access incurs high overhead on the critical path of program execution. On the other hand, page table (PT) scanning periodically checks the {\tt access} bit in all page table entries (PTE) to determine recently accessed pages since the last scanning. Compared to page fault-based tracking, which tracks every access on selected pages, PT scanning has to make a tradeoff between scanning overhead and tracking accuracy by choosing an appropriate scanning interval~\cite{memtis}. 


Linux adopts a {\em lazy} PT scanning mechanism to track hot pages, which lays the foundation for its tiered memory management. Linux maintains two LRU lists for a memory node: an {\em active} list to store hot pages and an {\em inactive} list for cold pages. By default, all new pages go to the inactive list and will be promoted to the active list according to two flags, \texttt{PG\_referenced} and \texttt{PG\_active}, in the per-page struct {\tt page}. {\tt PG\_reference} is set when the {\tt access} bit in the corresponding PTE is set upon a PTE check and {\tt PG\_active} is set after {\tt PG\_reference} is set for two consecutive times. A page is promoted to the active list when its {\tt PG\_active} flag is set. 
For file-backed pages, their accesses are handled by the OS through the file system interface, e.g., {\tt read()} and {\tt write()}. Therefore, their two flags are updated each time they are accessed. For anonymous pages, e.g., application memory allocated through {\tt malloc}, since page accesses are directly handled by the MMU hardware and bypass the OS kernel, the updates to their reference flags and LRU list management are only performed during memory reclamation. Under memory pressure, the swapping daemon {\tt kswapd} scans the inactive list and the corresponding PTEs to update inactive pages' flags, and reclaims/swaps out those with {\tt PG\_reference} unset. Additionally, {\tt kswapd} promotes hot pages (i.e., those with {\tt PG\_active} set) to the active list. This lazy scanning mechanism delays access tracking until it is necessary to reduce the tracking overhead, but undermines tracking accuracy.

TPP~\cite{Maruf_23asplos_tpp} leverages Linux's PT scanning to track hot pages and employs page fault-based tracking to decide whether to promote pages from slow memory. Specifically, TPP sets all pages residing in slow memory (e.g., CXL memory) as {\tt inaccessible}, and any user access to these pages will trigger a minor page fault, during which TPP decides whether to promote the faulting page. If the faulting page is on the active list, it is migrated (promoted) to the fast tier. Page demotion occurs when fast memory is under pressure and {\tt kswapd} migrates pages from the inactive list to slow memory. 

\begin{figure}[t]
	\centerline{\includegraphics[width=0.45\textwidth]{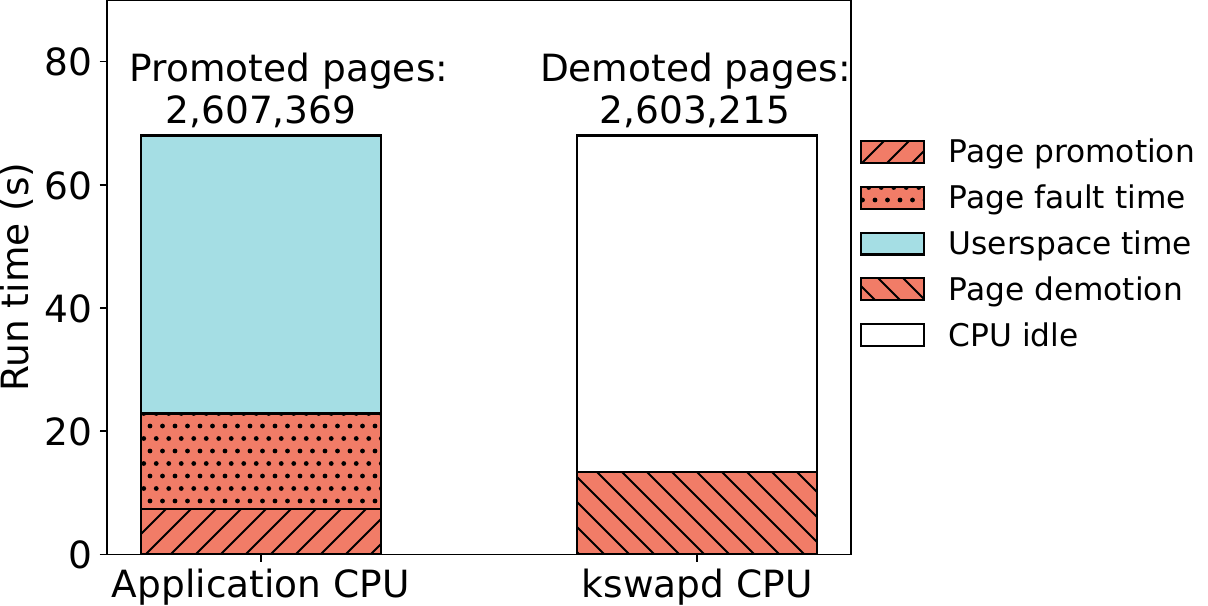}}
	\vspace{-1mm}
	\caption{\small Time breakdown in the execution of {\em TPP in progress}: Synchronous page migration and page fault handling account for a significant portion of the runtime.}
	\vspace{-4mm}
	\label{fig:TppVsNoMigrationPerformancebreakdown}
\end{figure}

Accurate and lightweight memory access tracking can be achieved with hardware support, e.g., by adding a PTE {\em count field} in hardware that records the number of memory accesses~\cite{7056027}. However, hardware-based tracking can increase the complexity and require extensive hardware changes in mainstream architectures (e.g., x86). In practice, the hardware-assisted sampling, such as via Processor Event-Based Sampling (PEBS)~\cite{memtis, Padmapriya_23asplos} on Intel platforms, has been employed to record page access (virtual address) information from sampled hardware events (e.g., LLC misses or store instructions). However, PEBS-based profiling also requires a careful balance between the frequency of sampling and the accuracy of profiling. We observed that the PEBS-based approach~\cite{memtis}, with a sampling rate optimized for minimizing overhead, remains coarse-grained and fails to capture many hot pages. Further, the sampling-based approach may not accurately measure access recency, thus limiting its ability to make timely migration decisions. 

\smallskip
\noindent{\bf Page migration} between memory tiers involves a complex procedure: \ding{172} The system must trap to the kernel (e.g., via page faults) to handle migration; \ding{173} The PTE of a migrating page must be locked to prevent others from accessing the page during migration and be \texttt{ummapped} from the page table; \ding{174} A translation lookaside buffer (TLB) shootdown must be issued to each processor (via inter-processor interrupts (IPIs)) that may have cached copies of the stale PTE; \ding{175} The content of the page is copied between tiers; \ding{176} Finally, the PTE must be \texttt{remapped} to point to the new location. Page migration can be done {\em synchronously} or {\em asynchronously}. Synchronous migration, e.g., page promotion in TPP, is on-demand triggered by user access to a page and on the critical path of program execution. During migration, the user program is blocked until migration is completed. Asynchronous migration, e.g., page demotion in TPP, is handled by a kernel thread (i.e., {\tt kswapd}), oftentimes off programs' critical path, when certain criteria are met. Synchronous migration is costly not only because pages are inaccessible during migration but also may involve a large number of page faults.

Figure~\ref{fig:TppVsNoMigrationPerformancebreakdown} shows the run time breakdown of the aforementioned benchmark while TPP is actively relocating pages between the two memory tiers. Since page promotion is synchronous, page fault handling and page content copying (i.e., promotion) are executed on the same CPU as the application thread. Page demotion is done through {\tt kswapd} and uses a different core. 
As shown in Figure~\ref{fig:TppVsNoMigrationPerformancebreakdown}, synchronous promotion together with page fault handling incurs significant overhead on the application core. In contrast, the demotion core remains largely idle and does not present a bottleneck. 
As will be discussed in Section~\ref{tpm}, userspace run time can also be prolonged due to repeated minor page faults (as many as 15) to successfully promote one page. This overhead analysis explains the poor performance of TPP observed in Figure~\ref{fig:TppVsNoMigrationPerformance}.


\subsection{Related Work}
A long line of pioneering work has explored a wide range of tiered storage/memory systems, built upon SSDs and HDDs~\cite{266253, 8714805, chen2011hystor, 5749736, 10.5555/1855511.1855519, 6557143, 5581609, 6005391, 267006},
DRAM and disks~\cite{270689, 270754, 269067, papagiannis2020optimizing}, HBM and DRAMs~\cite{7056027, batman, 6493624}, 
NUMA memory~\cite{autoNUMA, damon}, PM and DRAM~\cite{288741, abulila2019flatflash, Adnan_22hpca, thermostat}, local and far memory~\cite{Bergman_22ismm, 180194, 180727, 7266934, ruan2020aifm}, DRAM and CXL memory~\cite{li2023pond, Maruf_23asplos_tpp, memtis}, 
and multiple tiers~\cite{wu2021storage, Jonghyeon_21fast, Yan_19asplos_nimble, 288741, kwon2017strata}. We focus on tiered memory systems consisting of DRAM and the emerging byte-addressable memory devices, e.g., CXL memory and PM. \textsc{Nomad} also applies to other tiered memory systems such as HBM/DRAM and DRAM/PM.

\noindent{\bf Lightweight memory access tracking.} To mitigate software overhead associated with memory access tracking, Hotbox~\cite{Bergman_22ismm} employs two separate scanners for fast and slow tiers to scan the slow tier at a fixed rate while the fast tier at an adaptive rate, configurable based on the local memory pressure. Memtis~\cite{memtis} adjusts its PEBS-based sampling rate to ensure its overhead is under control (e.g., < 3\%). TMTS~\cite{Padmapriya_23asplos} also adopts a periodic scanning mechanism to detect frequency along with hardware sampling to more timely detect newly hot pages. While these approaches balance scanning/sampling overhead and tracking accuracy, an ``always-on'' profiling component does not seem practical, especially for high-pressure workloads. Instead, thermostat~\cite{thermostat} samples a small fraction of pages, while DAMON~\cite{damon} monitors memory access at a coarser-grained granularity (i.e., region). Although both can effectively reduce the scanning overhead, coarse granularity leads to lower accuracy regarding page access patterns. On the other hand, to reduce the overhead associated with frequent hint page faults like AutoNUMA~\cite{autoNUMA}, TPP~\cite{Maruf_23asplos_tpp} enables the page-fault based detection only for CXL memory (i.e., the slow tier) and tries to promote a page promptly via synchronous migration; prompt page promotion avoids subsequent page faults on the same page.

Inspired by existing lightweight tracking systems, such as Linux's active and inactive lists and hint page faults, \textsc{Nomad} advances them by incorporating more recency information with {\em no} additional CPU overhead. Unlike hardware-assisted approaches~\cite{memtis, Padmapriya_23asplos, Loh2012ChallengesIH}, \textsc{Nomad} does not require any additional hardware support.

\smallskip
\noindent{\bf Page migration optimizations.} To hide reclamation overhead from applications, TPP~\cite{Maruf_23asplos_tpp} decouples page allocation and reclamation; however, page migration remains in the critical path, incurring significant slowdowns. Nimble~\cite{Yan_19asplos_nimble} focuses on mitigating page migration overhead with  new migration mechanisms, including transparent huge page migration and concurrent multi-page migration. Memtis~\cite{memtis} further moves page migration out of the critical path using a kernel thread to promote/demote pages in the background. TMTS~\cite{Padmapriya_23asplos} leverage a user/kernel collaborative approach to control page migration. In contrast, \textsc{Nomad} aims to achieve prompt, on-demand page migration while moving page migration off the critical path. It is orthogonal to and can benefit from existing page migration optimizations. The most related work is \cite{concurrent}, which leverages hardware support to pin data in caches, enabling access to pages during migration. Again, \textsc{Nomad} does not need additional hardware support.  

%% file: design.tex
\section{\textsc{Nomad} Design and Implementation}
\label{design}
\textsc{Nomad} is a new page management mechanism for tiered memory that features {\em non-exclusive memory tiering} and {\em transactional page migration}. The goal of \textsc{Nomad} design is to enable the processor to freely access pages from both fast and slow memory tiers and move the cost of page migration off the critical path of users' data access. Note that \textsc{Nomad} does not make page migration decisions and relies on the existing memory access tracking in the OS to determine page temperature. Furthermore, \textsc{Nomad} does not impact the initial memory allocation in the OS and assumes a standard page placement policy. Pages are allocated from the fast tier whenever possible and are placed in the slower tier only when there is an insufficient number of free pages in the fast tier, or attempts to reclaim memory in the fast tier have failed. After the initial page placement, \textsc{Nomad} gradually migrates hot pages to the fast tier and cold pages to the slow tier. \textsc{Nomad} seeks to address two key issues: 1) {\em how to minimize the cost of page migration?} 2) {\em how to minimize the number of migrations?}

\smallskip
\noindent {\bf Overview}. Inspired by multi-level cache management in modern processors, which do not employ a purely inclusive or exclusive caching policy between tiers~\cite{Alian-micro21} to facilitate the sharing of or avoid the eviction of certain cache lines, \textsc{Nomad} embraces a {\em non-exclusive} memory tiering policy to prevent memory thrashing when under memory pressure. Unlike the existing page management schemes that move pages between tiers and require that a page is only present in one tier, \textsc{Nomad} instead copies pages from the slow tier to the fast tier and keeps a shadow copy of the migrated pages at the slow tier. The non-exclusive tiering policy maintains shadow copies only for pages that have been promoted to the fast tier, thereby not an inclusive policy. The advantage of the non-exclusive policy is that the demotion of clean, cold pages can be simplified to remapping the page table entry (PTE) without the need to copy the cold page to the slower tier. 

The building block of \textsc{Nomad} is a new {\em transactional page migration} (TPM) mechanism to reduce the cost of page migrations. Unlike the existing unmap-copy-remap 3-step page migration, TPM opportunistically copies a page without unmapping it from the page table. During the page copy, the page is not locked and can be accessed by a user program. After the copy is completed, TPM checks if the page has been dirtied during the copy. If not, TPM locks the page and remaps it in the PTE to the faster tier. Otherwise, the migration is aborted and will be tried at a later time. TPM not only minimizes the duration during which a page is inaccessible but also makes page migration asynchronous, thereby removing it from the critical path of users' data access. 

Without loss of generality, we describe \textsc{Nomad} design in the context of Linux. 
We start with transactional page migration and then delve into page shadowing -- an essential mechanism that enables non-exclusive memory tiering.

\begin{figure}[t]
	\centerline{\includegraphics[width=0.46\textwidth]{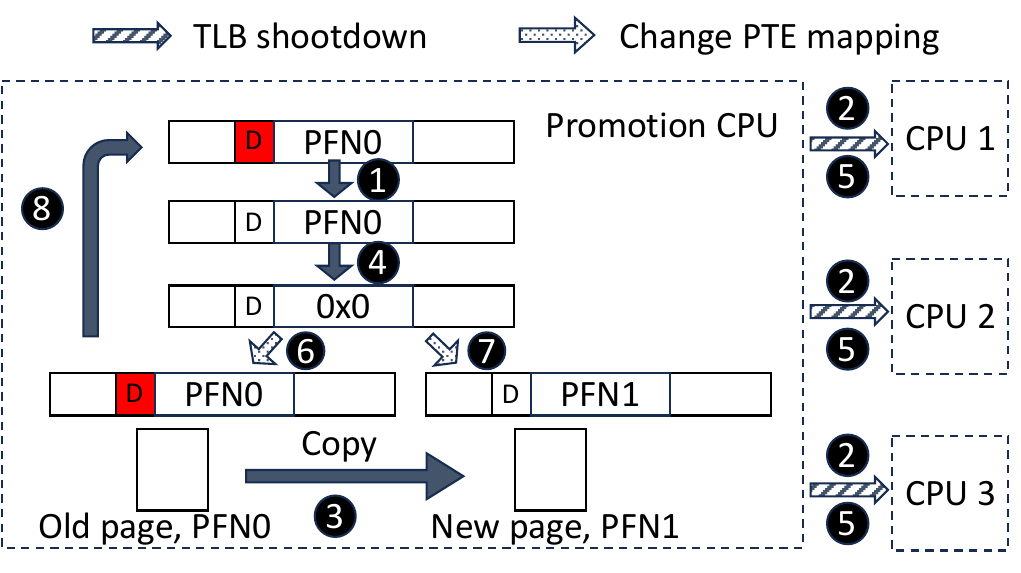}}
	\vspace{-1mm}
	\caption{The workflow of transactional page migration. PFN is the page frame number and D is the dirty bit in PTE. The page is only inaccessible by user programs during step 4 when the page is remapped in the page table.}
	\vspace{-4mm}
	\label{fig:tpm}
\end{figure}
\subsection{Transactional Page Migration}
\label{tpm}
The motivation to develop {\small TPM} is to make page migration entirely asynchronous and decoupled from users' access to the page. As discussed in Section~\ref{sec:page-management}, the current page migration in Linux is synchronous and on the critical path of users' data access. For example, the default tiered memory management in Linux, TPP, attempts to migrate a page from the slow tier whenever a user program accesses the page. Since the page is in {\tt inaccessible} mode, the access triggers a minor page fault, leading TPP to attempt the migration. The user program is blocked and makes no progress until the minor page fault is handled and the page is remapped to the fast tier, which can be a time-consuming process. Worse, if the migration fails, the OS remains in function {\tt migrate\_pages} and retries the aforementioned migration until it is successful or reaching a maximum of 10 attempts. 

TPM decouples page migration from the critical path of user programs by making the migrating page accessible during migration. Therefore, users will access the migrating page from the slow tier before the migration is complete. While accessing a hot page from the slow tier may lead to sub-optimal memory performance, it avoids blocking user access due to the migration, thereby leading to superior user-perceived performance. Figure~\ref{fig:tpm} shows the workflow of TPM. Before migration commences, TPM clears the protection bit of the page frame and adds the page to a migration pending queue. Since the page is no longer protected and not yet unmapped from the page table, following accesses to the page will not trigger additional page faults. 

TPM starts a migration transaction by clearing the dirty bit of the page (step \ding{182}) and checks the dirty bit after the page is copied to the fast tier to determine whether the transaction was successful. After changing the dirty bit in PTE, TPM issues a TLB shootdown to all cores that ever accessed this page (step \ding{183}). This is to ensure that subsequent writes to the page can be recorded on the PTE. After the TLB shootdown is completed, TPM starts copying the page from the slow tier to the fast tier (step~\ding{184}).  To commit the transaction, TPM checks the dirty bit by loading the entire PTE using atomic instruction {\tt get\_and\_clear} (step \ding{185}). Clearing the PTE is equivalent to unmapping the page and thus another TLB shootdown is needed (step \ding{186}). Note that after unmapping the page from PTE, it becomes inaccessible by users. TPM checks whether the page was dirtied during the page copy (step \ding{187}) and either commits the transaction by remapping the page to the fast tier if the page is clean (step \ding{188}) or otherwise aborts the transaction (step \ding{189}). If the migration is aborted, the original PTE is restored and waits for the next time when TPM is rescheduled to retry the migration. The duration in which the page is inaccessible is between~\ding{185} and~\ding{188}/~\ding{189}, significantly shorter than that in TPP (possibly multiple attempts between ~\ding{182} and~\ding{188}).

Page migration is a complex procedure that involves memory tracing and updates to the page table for page remapping. The state-of-the-art page fault-based migration approaches, e.g., TPP in Linux~\cite{Maruf_23asplos_tpp}, employ synchronous page migration, a mechanism in the Linux kernel for moving pages between NUMA nodes. In addition to the extended migration time affecting the critical path of user programs, this mechanism causes excessive page faults when integrated with the existing LRU-based memory tracing. TPP makes {\em per-page} migration decisions based on whether the page is on the {\em active} LRU list. Nevertheless, in Linux, memory tracing adds pages from the inactive to the active LRU list in batches of 15 requests~\footnote{The 15 requests could be repeated requests for promoting the same page to the active LRU list}, aiming to minimize the queue management overhead. Due to synchronous page migration, TPP may submit multiple requests (up to 15 if the request queue is empty) for a page to be promoted to the active LRU list to initiate the migration process. In the worst case, migrating one page may generate as many as 15 minor page faults.   

\begin{figure}[t]
	\centerline{\includegraphics[width=0.38\textwidth]{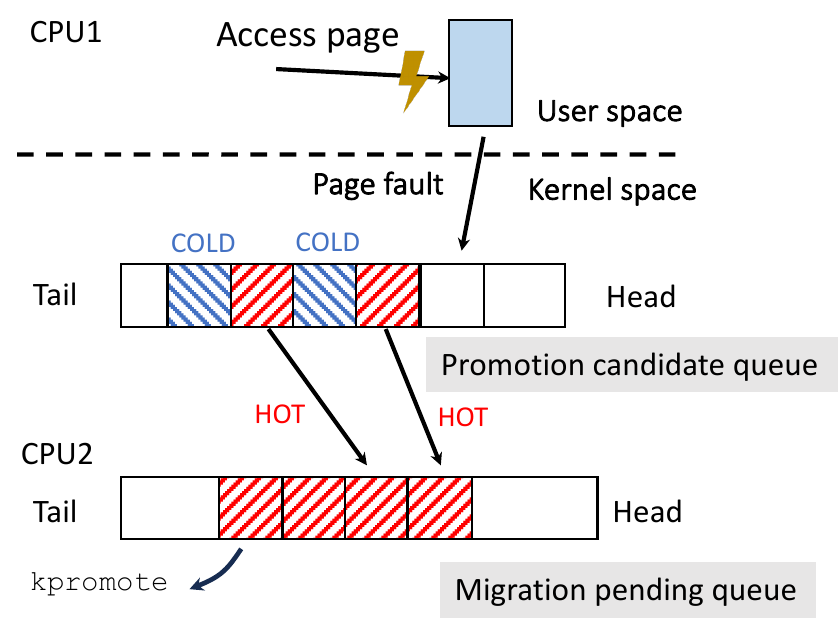}}
	\vspace{-1mm}
	\caption{TPM uses a two-queue design to enable asynchronous page migration.}
	\vspace{-4mm}
	\label{fig:LookbackQueue}
\end{figure}

TPM provides a mechanism to enable {\bf asynchronous page migration} but requires additional effort to interface with memory tracing in Linux to minimize the number of page faults needed for page migration. As shown in Figure~\ref{fig:LookbackQueue}, in addition to the inactive and active LRU lists in memory tracing, TPM maintains a separate {\em promotion candidate} queue (PCQ) for pages that 1) have been tried for migration but 2) not yet promoted to the active LRU list. Upon each time a minor (hint) page fault occurs and the faulting page is added to PCQ, TPM checks if there are any hot pages in PCQ that have both the {\tt active} and {\tt accessed} bits set. These hot pages are then inserted to a {\em migration pending queue}, from where they will be tried for asynchronous, transactional migration by a background kernel thread {\tt kpromote}. Note that TPM does not change how Linux determines the temperature of a page. For example, in Linux, all pages in the active LRU list, which are eligible for migration, have the two memory tracing bits set. However, not all pages with these bits set are in the active list due to LRU list management. TPM bypasses the LRU list management and provides a more efficient method to initiate page migration. If all transactional migrations were successful, TPM guarantees that only one page fault is needed per migration in the presence of LRU list management.          

\subsection{Page Shadowing}
\label{shadowing}

\begin{figure}
	\centerline{\includegraphics[width=0.4\textwidth]{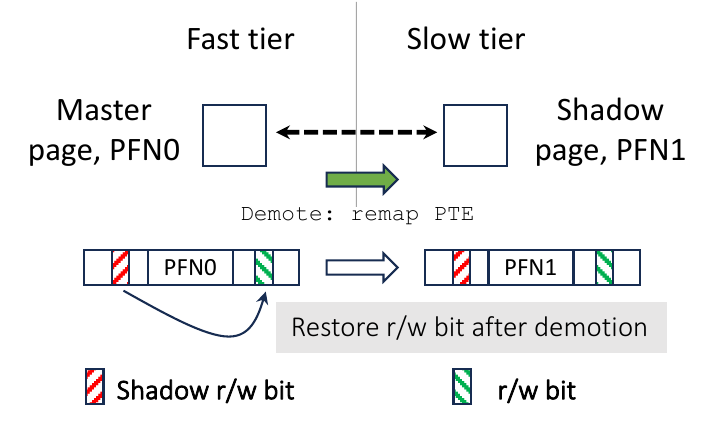}}
	\vspace{-1mm}
	\caption{Shadow page management using {\tt shadow r/w} bit.}
	\vspace{-4mm}
	\label{fig:shadow}
\end{figure}

To enable non-exclusive memory tiering, \textsc{Nomad} introduces a one-way {\em page shadowing} mechanism to allow a subset of pages resident in the performance tier to have a shadow copy in the capacity tier. Only pages promoted from the slow tier have shadow copies in the slow tier. Shadow copies are the original pages residing on the slow tier before they are unmapped in the page table and migrated to the fast tier. Shadow pages play a crucial role in minimizing the overhead of page migration during periods of memory pressure. Instead of swapping hot and cold pages between memory tiers, page shadowing enables efficient page demotion through page table remapping. This would eliminate half of the page migration overhead, i.e., page demotion, during memory thrashing. 

\smallskip
\noindent {\bf Indexing shadow pages}. Inspired by the indexing of file-based data in the Linux page cache, \textsc{Nomad} builds an XArray for indexing shadow pages. An XArray is a radix-tree like, cache-efficient data structure that acts as a key-value store, mapping from the physical address of a fast tier page to the physical address of its shadow copy on the slow tier. 
Upon successfully completing a page migration, \textsc{Nomad} inserts the addresses of both the new and old pages into the XArray. Additionally, it adds a new {\tt shadow} flag to the {\tt struct page} of the new page, indicating that shadowing is on for this page.

\smallskip
\noindent {\bf Shadow page management}. The purpose of maintaining shadow pages is to assist with page demotion. Fast or efficient page demotion is possible via page remapping if the master page, i.e., the one on the fast tier, is clean and consistent with the shadow copy. Otherwise, the shadow copy should be discarded. To track inconsistency between the master and shadow copies, \textsc{Nomad} sets the master page as {\tt read-only} and a write to the page causes a page fault. To simplify system design and avoid additional cross-tier traffic, \textsc{Nomad} discards the shadow page if the master page is dirtied. 

However, tracking updates to the master page poses a significant challenge. Page management in Linux relies heavily on the read-write permission to perform various operations on a page, such as copy-on-write (CoW). While setting master pages as {\tt read-only} effectively captures all writes, it may affect how these master pages are managed in the kernel. To address this issue, \textsc{Nomad} introduces a procedure called {\em shadow page fault}. 
It still designates all master pages as {\tt read-only} but preserves the original read-write permission in an unused software bit on the page's PTE (as shown in Figure~\ref{fig:shadow}). We refer to this software bit as {\tt shadow r/w}. Upon a write to a master page, a page fault occurs. Unlike an ordinary page fault that handles write violation, the shadow page fault, which is invoked if the page's {\tt shadow} flag is set in its {\tt struct page}, restores the read-write permission of the faulting page according to the {\tt shadow r/w} bit and discards/frees the shadow page. The write may proceed once the shadow page fault returns and reinstates the page to be writable. For {\tt read-only} pages, tracking shadow pages does not impose additional overhead; for writable pages, it requires one additional shadow page fault to restore their write permission.  

\smallskip
\noindent {\bf Reclaiming shadow pages}. Non-exclusive memory tiering introduces space overhead due to the storage of shadow pages. If shadow pages are not timely reclaimed when the system is under memory pressure, applications may encounter out-of-memory (OOM) errors, which would not occur under exclusive memory tiering. There are two scenarios in which shadow pages should be reclaimed. First, the Linux kernel periodically checks the availability of free pages and if free memory falls below {\tt low\_water\_mark}, kernel daemon {\tt kswapd} is invoked to reclaim memory. \textsc{Nomad} instructs {\tt kswapd} to prioritize the reclamation of shadow pages. Second, upon a memory allocation failure, \textsc{Nomad} also tries to free shadow pages. To avoid OOM errors, the number of freed shadow pages should exceed the number of requested pages. However, frequent memory allocation failures could negatively affect system performance. 
\textsc{Nomad} employs a straightforward heuristic to reclaim shadow pages, targeting 10 times the number of requested pages or until all shadow pages are freed. 
While excessive reclamation may have a negative impact on \textsc{Nomad}'s performance, it is crucial to prevent Out-of-Memory (OOM) errors. Experiments in Section~\ref{sec:eval} demonstrate the robustness of \textsc{Nomad} even under extreme circumstances.

\subsection{Limitations}
\label{sec:limitation}
\textsc{Nomad} relies on two rounds of TLB shootdown to effectively track updates to a migrating page during transactional page migration. When a page is used by multiple processes or mapped by multiple page tables, its migration involves multiple TLB shootdowns, per each mapping, that need to happen simultaneously. The overhead of handling multiple IPIs could outweigh the benefit of asynchronous page copy. Hence, \textsc{Nomad} deactivates transactional page migration for multi-mapped pages and resorts to the default synchronous page migration mechanism in Linux. As high-latency TLB shootdowns based on IPIs continue to be a performance concern, modern processors, such as ARM, future AMD, and Intel x86 processors, are equipped with ISA extensions for faster broadcast-based~\cite{arm_tlb_broadcast, amd_tlb_broadcast} or micro-coded RPC-like~\cite{intelcpurpc} TLB shootdowns. These emerging lightweight TLB shootdown methods will greatly reduce the overhead of TLB coherence in tiered memory systems with expanded memory capacity. \textsc{Nomad} will also benefit from the emerging hardware and can be extended to scenarios where more intensive TLB shootdowns are necessary. 


%% file: evaluation.tex
\section{Evaluation}
\label{sec:eval}

\begin{figure}[t]
	\centerline{\includegraphics[width=0.35\textwidth]{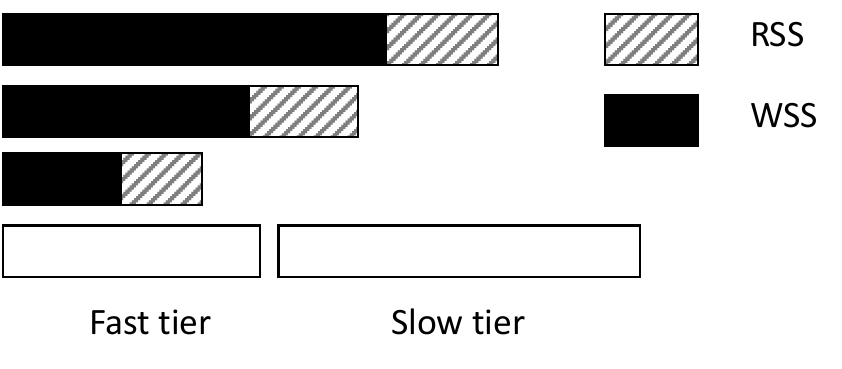}}
	\vspace{-1mm}
	\caption{The three memory provisioning schemes used in the evaluation. From bottom to top concerning fast memory: over-provisioning, approaching capacity, and under-provisioning.}
        \label{fig:scenarios}
	\vspace{-4mm}
\end{figure}

This section presents a thorough evaluation of \textsc{Nomad}, focusing on its performance, overhead, and robustness. Our primary goal is to understand tiered memory management by comparing \textsc{Nomad} with existing representative approaches to reveal the benefits and potential limitations of current page management approaches for emerging tiered memory. 

We analyze two types of memory footprints: 1) resident set size (RSS) – the total size of memory occupied by a program, and 2) working set size (WSS) – the amount of memory a program actively uses during execution. RSS determines the initial page placement, while WSS dictates the number of pages that should be migrated to the fast tier. Since we focus on in-memory computing, WSS is typically smaller than RSS. Figure~\ref{fig:scenarios} illustrates the three scenarios we study with the WSS size smaller than, close to, and larger than fast memory size.

\begin{table*}[t]
\centering
\footnotesize
\begin{tabular}{ c | c | c | c | c }
  \hline

  \thead{} & Platform A  & Platform B & Platform C & Platform D\\
  \thead{} &   & (engineering sample) &   &   \\
  \hline
  \hline
  CPU & \makecell{4th Gen  Xeon Gold \\ 2.1GHz} & \makecell{4th Gen  Xeon Platinum \\ 3.5GHz} & \makecell{2nd Gen  Xeon Gold \\ 3.9GHz}  & \makecell{AMD Genoa \\ 3.7GHz } \\
  \hline
  Performance tier (DRAM) & \makecell{16 GB  DDR5} & \makecell{16 GB   DDR5} & \makecell{16 GB  DDR4}  &  \makecell{16GB DDR5} \\
  \hline
    \makecell{Capacity tier \\ (CXL or PM  Memory) } & \makecell{Agilex 7 \\  16 GB   DDR4} & \makecell{Agilex 7 \\  16 GB   DDR4} & \makecell{Optane 100 \\ 256 GB DDR-T $\times$ 6}  & \makecell{Micron CXL memory \\ 256GB $\times$ 4} \\
  \hline
 Performance tier  read latency & \makecell{316 cycles} & \makecell{226 cycles} & \makecell{249 cycles}  &  \makecell{391 cycles} \\
  \hline
 Capacity tier  read latency & \makecell{854 cycles} & \makecell{737 cycles} & \makecell{1077 cycles}  &  \makecell{712 cycles} \\
  \hline
 \makecell{Performance tier \\  bandwidth (GB/s) \\ Single Thread / Peak performance} & \makecell{Read: 12/31.45 \\ Write: 20.8/28.5} & \makecell{Read: 12/31.2 \\ Write: 22.3/23.67} & \makecell{Read: 12.57/116 \\ Write: 8.67/85}  &  \makecell{Read: 37.8/270 \\ Write: 89.8/272} \\
  \hline
   \makecell{Capacity tier \\ bandwidth (GB/s) \\ Single Thread/Peak performance} & \makecell{Read: 4.5/21.7 \\ Write: 20.7/21.3} & \makecell{Read: 4.45/22.3 \\ Write: 22.3/22.4} & \makecell{Read: 4/40.1 \\ Write: 8.1/13.6}  &  \makecell{Read: 20.25/83.2 \\ Write: 57.7/84.3} \\
  \hline
\end{tabular}
\caption{The configurations of four testbeds and performance characteristics of various memory devices.  }
\label{tab:config}
\end{table*}

\begin{table*}[t]
\centering
\footnotesize
  
  \begin{tabular}{c|c|c|c|c}
    \hline
    Workload Type & In progress Promotion & In progress Demotion & Steady Promotion & Steady demotion \\
    \hline
    \hline
    Small WSS & \makecell{(1.2M|1M)/(15.9K|134K)/ \\ (1.16M|781K)}    & \makecell{(2.4M|2.2M)/(15.9K|140K)/ \\ (2.7M|1.5M)}  & \makecell{(0|3.3K)/(7.7K|104K)/ \\ (82|74)}    & \makecell{(424K|56K)/(0|104K)/ \\ (48K|0)}   \\
     \hline
    Medium WSS &  \makecell{(4M|6M)/(0|0)/ \\ (1.6M|5M)}   & \makecell{(4.7M|6M)/(2|512)/ \\ (2.5M|4.8M)}   & \makecell{(1.8M|3.2M)/(17.4K|0)/ \\ (417K|1.6M)}   & \makecell{(1.9M|3.2M)/(16.9K|0)/ \\ (293K|1.4M)}   \\
     \hline
  Large WSS  & \makecell{(7M|5.9M)/(0|0)/ \\ (4.5M|7M)}   & \makecell{(7.2M|6.5M)/(0|15)/ \\ (4.1M|7.2M)}   & \makecell{(7.1M|5.2M)/(0|143K)/ \\ (6.8M|8.8M)}   & \makecell{(7.1M|5.3M)/(0|143K)/ \\ (6.8M|8.9M)}  \\
   \hline
\end{tabular}
  \caption{
  The number of page promotions/demotions for read|write during the {\em migration in progress} and the {\em stable} phases for TPP/Memtis-Default/NOMAD. The data corresponds to Figure \ref{fig:cam-ready-microbench-bw-platform-a} for platform A. 
  }
  \vspace{-4mm}
  \label{tab:cam-ready-pagemigrationnumb}
\end{table*}

\smallskip
\noindent{\bf Testbeds.} We conducted experiments on {\em four} platforms with different configurations in CPU, local DRAM, CXL memory, and persistent memory, as detailed in Table~\ref{tab:config}. 
\begin{itemize}[leftmargin=*]
\setlength{\itemindent}{0em}
\setlength\itemsep{0em}
\item {\em Platform A} was built with commercial off-the-shelf (COTS) Intel Sapphire Rapids processors and a 16 GB Agilex-7 FPGA-based CXL memory device~\cite{agilex}. 

\item {\em Platform B} featured an engineering sample of the Intel Sapphire Rapids processors with the same FPGA-based CXL memory device. The prototype processors have engineering tweaks that have the potential to enhance the performance of CXL memory, which were not available on platform A. 

\item {\em Platform C} included an Intel Cascade Lake processor and six 256 GB 100 series Intel Optane Persistent Memory. This platform enabled the full capability of PEBS-based memory tracking and allowed for a comprehensive comparison between page fault- and sampling-based page migration.

\item {\em Platform D} had an AMD Genoa 9634 processor and four 256 GB Micron's (pre-market) CXL memory modules. This platform allowed us to evaluate \textsc{Nomad} with more realistic CXL memory configurations.
\end{itemize}

Since the FPGA-based CXL memory device had only 16 GB of memory, we configured local DRAM to 16 GB for all platforms~\footnote{Although platform C and D have larger PM or CXL memory sizes, we configured them with 16 GB slow memory consistent with platform A and B for a fair comparison in micro-benchmarks.This limit was lifted when testing real applications.}. Note that platform C was equipped with DDR4 DRAM as fast memory while the other platforms used DDR5 DRAM. We evaluated both CXL memory and persistent memory (PM) as slow memory. Table~\ref{tab:config} lists the performance characteristics of the four platforms for single-threaded and peak (multi-threaded) performance. While CXL memory and PM have distinct characteristics, including persistence, concurrent performance, and read/write asymmetry, they achieve comparable performance within 2-3x of DRAM and provide a similar programming interface as a CPUless memory node. To ensure a fair comparison, we only enabled one socket on each of the four platforms. Intel platforms were configured with 32 cores while the AMD platform had 84 cores.

\smallskip

\noindent{\bf Baselines for comparison.} We compared \textsc{Nomad} with two state-of-the-art tired memory systems: TPP~\cite{Maruf_23asplos_tpp} and Memtis~\cite{memtis}. We evaluated both TPP and Nomad on Linux kernel v5.13-rc6 and ran Memtis on kernel v5.15.19, the kernel version upon which Memtis was built and released. We tested two versions of Memtis -- Memtis-Default and Memtis-QuickCool -- with different data cooling speeds (i.e., the number of samples collected before halving a page's access count). Specifically, Memtis-Default used the default cooling period of 2,000k samples, while Memtis-QuickCool used a period of 2k samples. A shorter cooling period encourages more frequent page migration between the memory tiers.

\begin{figure*}[h!]
    \centering
    \begin{minipage}[b]{0.33\textwidth}
        \centering
        \includegraphics[width=\textwidth]{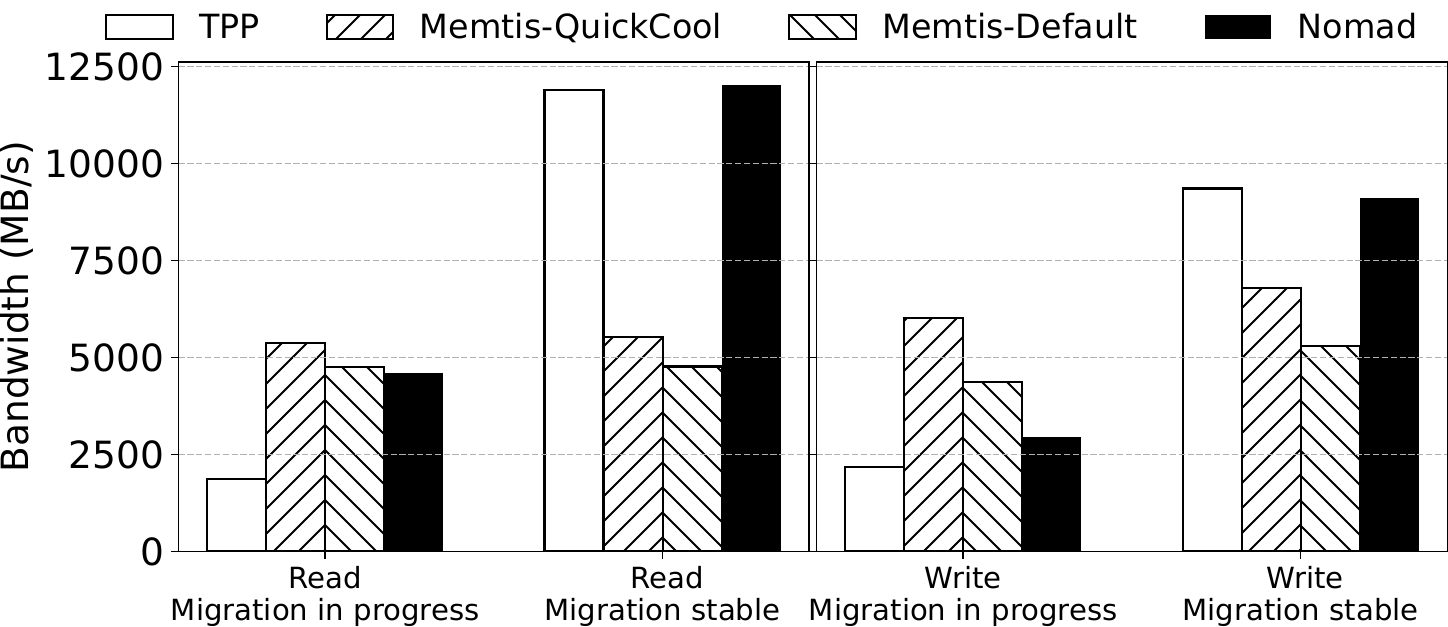}
        \caption*{(a) Small working set size}
    \end{minipage}
    \hfill
    \begin{minipage}[b]{0.33\textwidth}
        \centering
        \includegraphics[width=\textwidth]{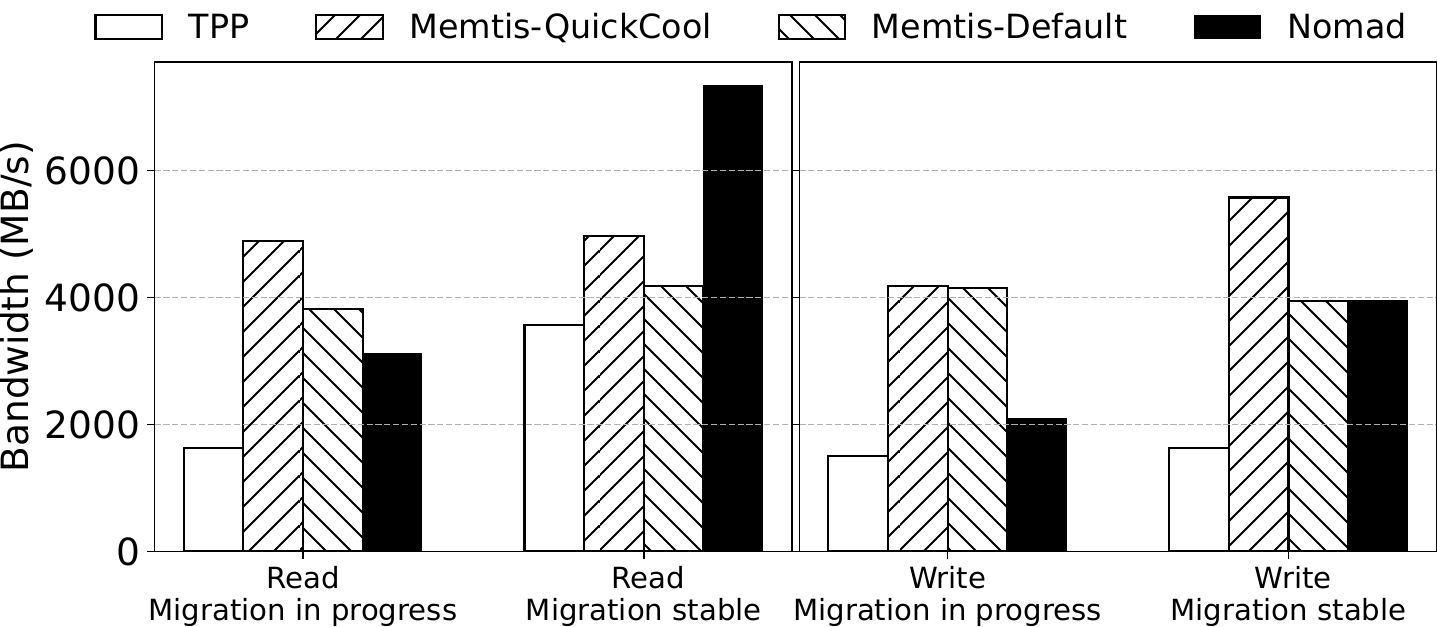}
        \caption*{(b) Medium working set size}
    \end{minipage}
    \hfill
    \begin{minipage}[b]{0.33\textwidth}
        \centering
        \includegraphics[width=\textwidth]{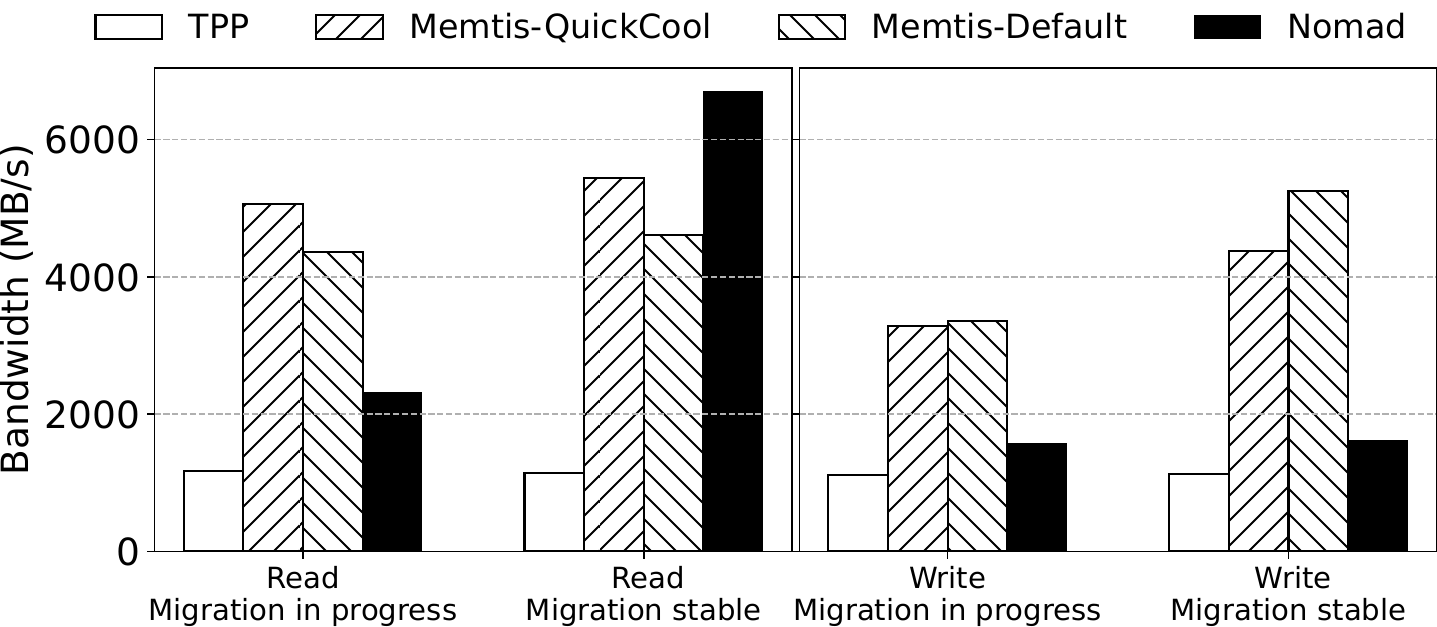}
        \caption*{(c) Large working set size}
    \end{minipage}
    \caption{Performance comparison between TPP, Memtis-Default, Memtis-QuickCool, and \textsc{Nomad} on platform A.}
    \vspace{-4mm}
    \label{fig:cam-ready-microbench-bw-platform-a}
\end{figure*}

Memtis relies on Intel's Processor Event-Based Sampling (PEBS) to track memory access patterns. It samples various hardware events, including LLC misses, TLB misses, and retired store instructions, to infer accessed page addresses and build frequency-based histograms to aid in making migration decisions. Memtis currently only supports Intel-based systems, though it can be ported to AMD processors with Instruction-based Sampling (IBS). Thus, Memtis was not evaluated on platform D. Memtis works slightly differently on CXL-memory systems (platforms A and B) and the PM system (platform C). LLC misses to CXL memory are regarded as {\em uncore} events on Intel platforms and thus cannot be captured by PEBS. Therefore, Memtis relies solely on TLB misses and retired store instructions to infer page temperature on platforms A and B.

\subsection{Micro-benchmarks}
To evaluate the performance of \textsc{Nomad}'s transactional page migration and shadowing mechanisms, we developed a micro-benchmark to precisely assess \textsc{Nomad} in a controlled manner. This micro-benchmark involves 1) allocating data to specific segments of the tiered memory; 2) running tests with various working set sizes (WSS) and resident set sizes (RSS); and 3) generating memory accesses to the WSS data that mimic real-world memory access patterns with a Zipfian distribution. We created three scenarios representing small, medium, and large WSS, as illustrated in Figure~\ref{fig:scenarios}, to evaluate tiered memory management under different memory pressures. 
As platform B behaved similarly to platform A in micro-benchmarks, it is excluded from the discussion.

\smallskip
\noindent{\bf Small WSS.} We began with a scenario with a small WSS of 10 GB and a total RSS of 20 GB. Initially, we filled the first 10 GB of local DRAM with the first half of the RSS data. Subsequently, we allocated 10 GB of WSS data as the second half of the RSS -- 6 GB on the local DRAM and 4 GB on CXL memory (platforms A and D) and the PM (platform C). 
The micro-benchmark continuously performed memory reads or writes (following a Zipfian distribution) to this 10 GB WSS data, spread across both the local DRAM and CXL memory or PM. The frequently accessed, or ``hot'' data, was uniformly distributed along the 10 GB WSS.  In TPP and \textsc{Nomad}, accessing data on CXL memory or PM triggered page migration to the local DRAM, with TPP performing this migration synchronously and \textsc{Nomad} asynchronously. In contrast, Memtis used a background thread to migrate hot data from CXL memory or PM to the local DRAM. Due to page migration, the 4 GB WSS data, initially allocated to CXL memory or PM, was gradually moved to the local DRAM. Since the WSS was small (i.e., 10 GB), it could be completely stored in the fast tier (i.e., local DRAM) after the micro-benchmark reached a {\em stable} state.

\begin{figure*}[t!]
    \centering
    \begin{minipage}[b]{0.33\textwidth}
        \centering
        \includegraphics[width=\textwidth]{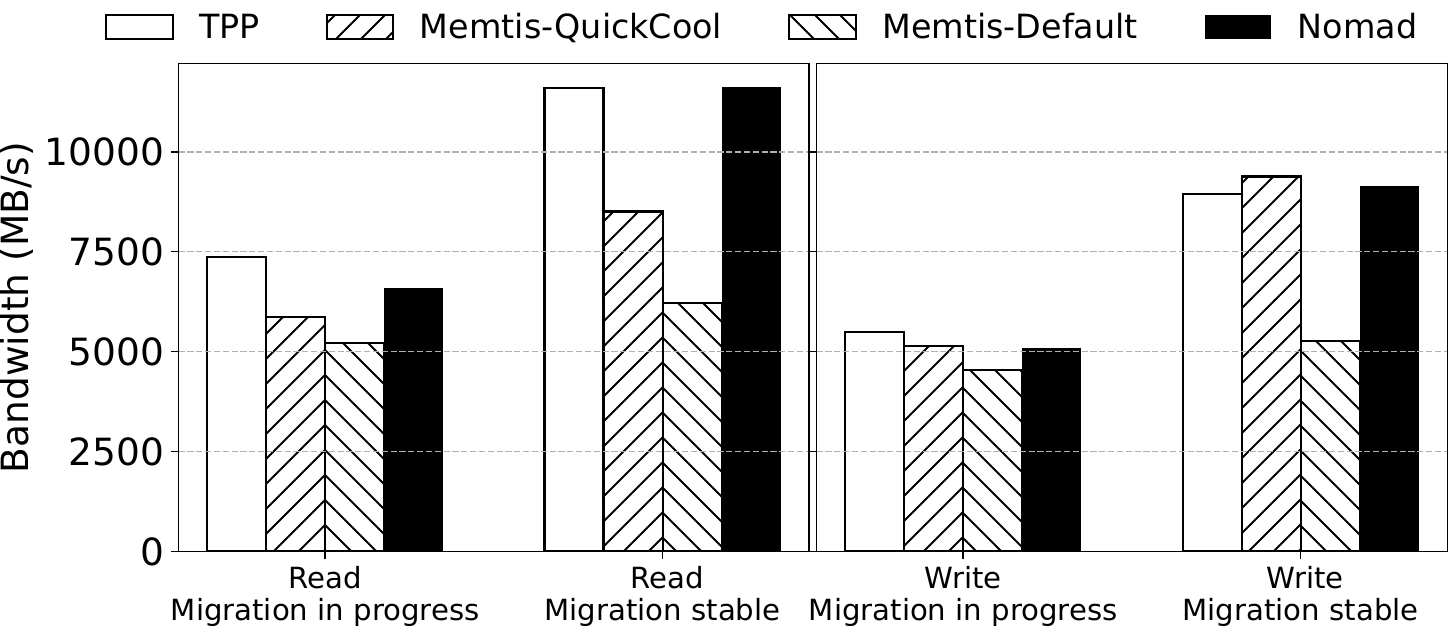}
        \caption*{(a) Small working set size}
    \end{minipage}
    \hfill
    \begin{minipage}[b]{0.33\textwidth}
        \centering
        \includegraphics[width=\textwidth]{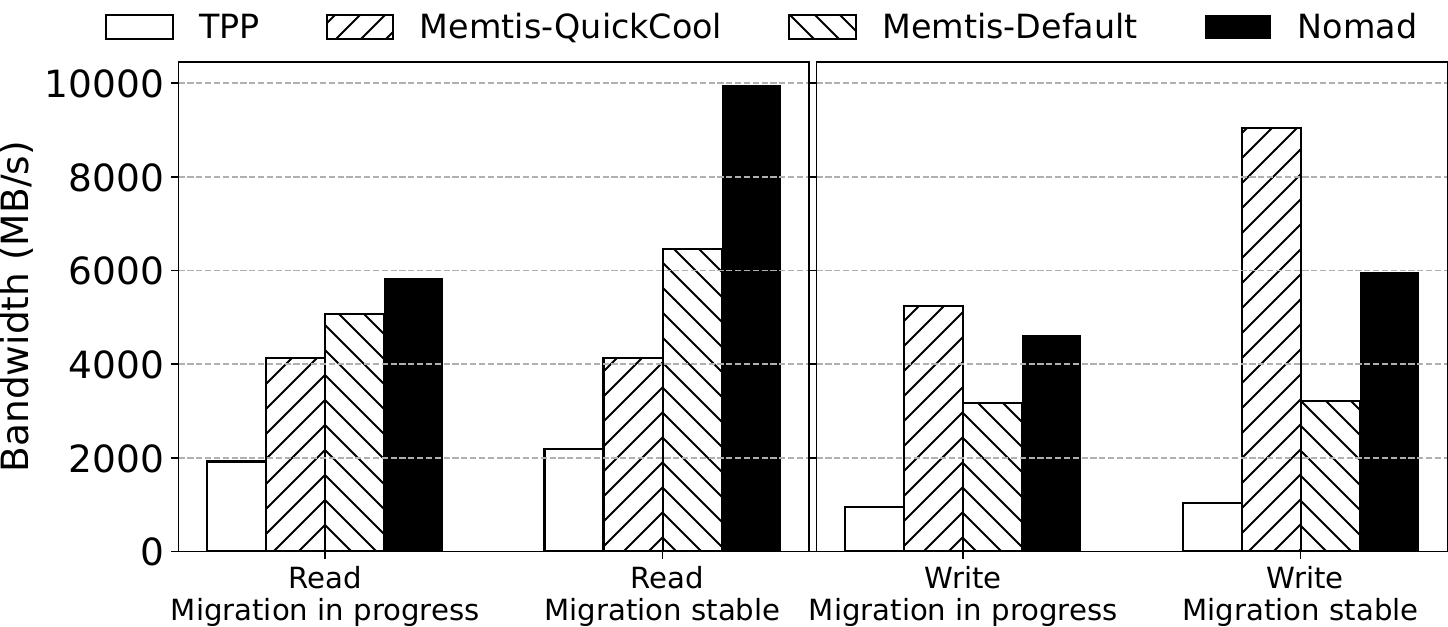}
        \caption*{(b) Medium working set size.}
    \end{minipage}
    \hfill
    \begin{minipage}[b]{0.33\textwidth}
        \centering
        \includegraphics[width=\textwidth]{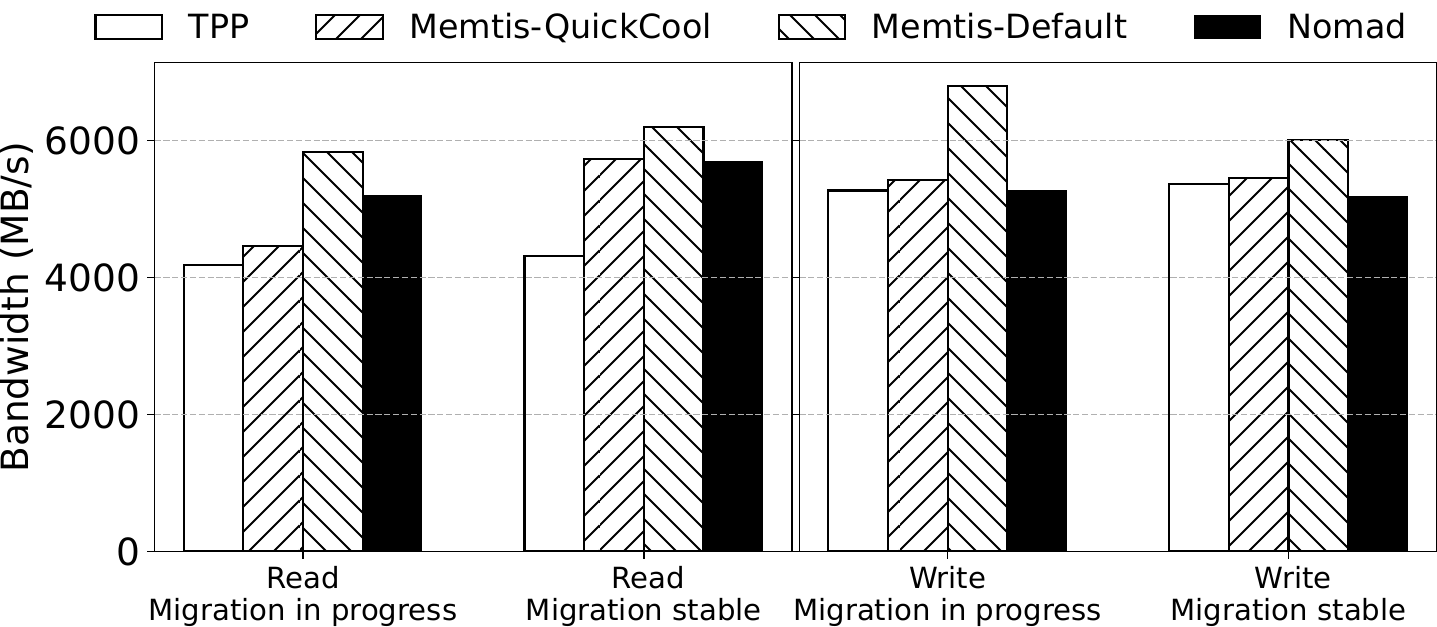}
        \caption*{(c) Large working set size}
    \end{minipage}
    \caption{Performance comparison between TPP, Memtis-Default, Memtis-QuickCool, and \textsc{Nomad} on platform C.}
    \label{fig:cam-ready-microbench-bw-platform-c}
\end{figure*}

\begin{figure*}[h!]
    \centering
    \begin{minipage}[b]{0.3\textwidth}
        \centering
        \includegraphics[width=\textwidth]{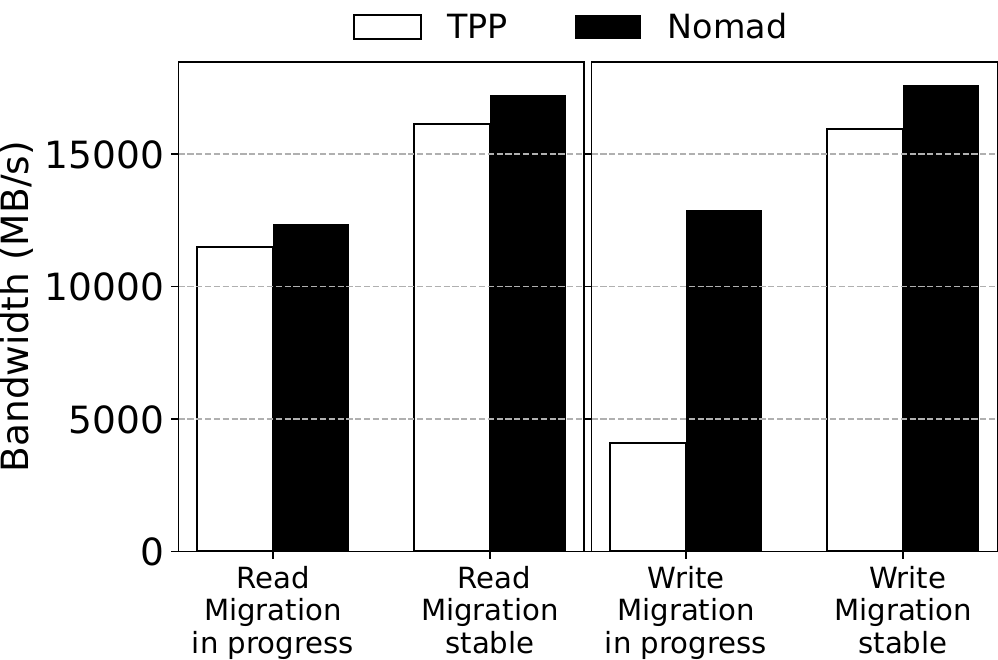}
        \caption*{(a) Small working set size}
    \end{minipage}
    \hfill
    \begin{minipage}[b]{0.3\textwidth}
        \centering
        \includegraphics[width=\textwidth]{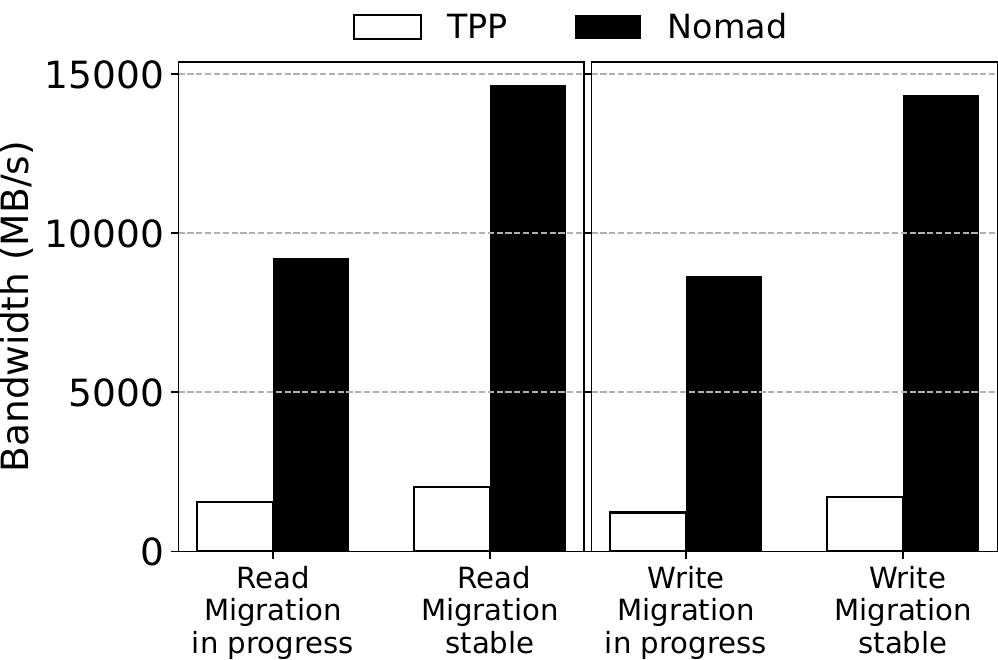}
        \caption*{(b) Medium working set size}
    \end{minipage}
    \hfill
    \begin{minipage}[b]{0.3\textwidth}
        \centering
        \includegraphics[width=\textwidth]{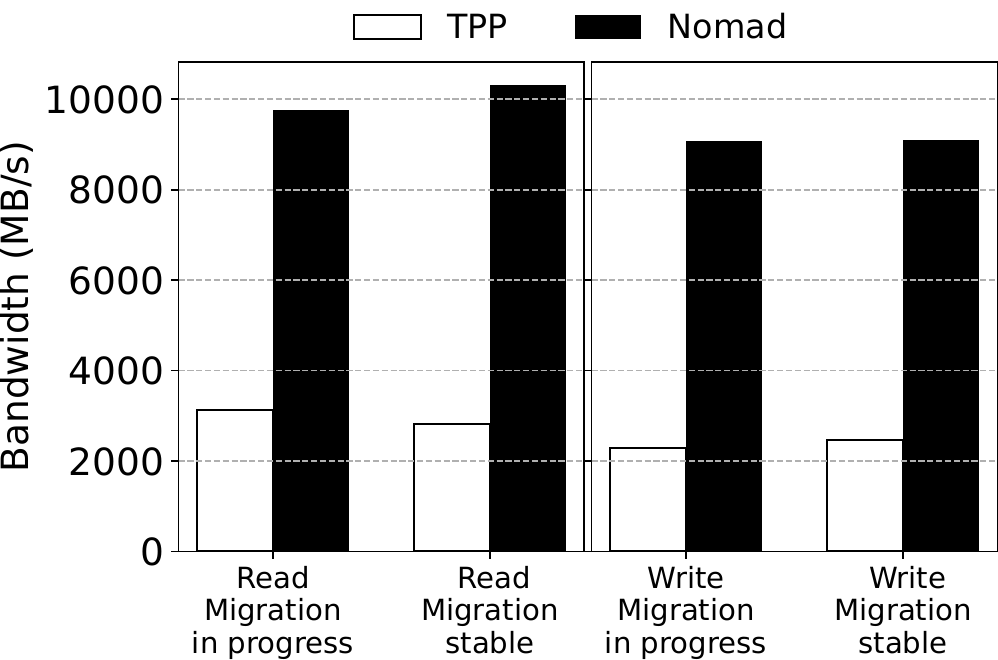}
        \caption*{(c) Large working set size}
    \end{minipage}
    \caption{Performance comparison between TPP and \textsc{Nomad} on platform D with AMD Genoa processor. Memtis does not support AMD's instruction-based sampling (IBS) and thus was not tested.
    }
    \vspace{-4mm}
    \label{fig:cam-ready-microbench-bw-platform-d}
\end{figure*}

Figures~\ref{fig:cam-ready-microbench-bw-platform-a} (a),~\ref{fig:cam-ready-microbench-bw-platform-c} (a), and~\ref{fig:cam-ready-microbench-bw-platform-d} (a) show that in the {\em transient} phase, during which page migration was conducted intensively (i.e., {\em migration in progress}), both \textsc{Nomad} and Memtis demonstrated similar performance regarding memory bandwidth for reads. Although page-fault-based page migration in \textsc{Nomad} could incur more overhead than the PEBS-based approach in Memtis, when the WSS can fit in fast memory and no memory thrashing occurs, the benefit of migration outweighs its overhead. 
For writes, e.g., on platform A, \textsc{Nomad} incurred noticeable performance degradation compared to Memtis due to possibly aborted migrations and the maintenance of shadow pages. Note that \textsc{Nomad}'s overhead varies across platforms depending on the performance difference between fast and slow memory. 
In contrast, 
\textsc{Nomad}
consistently 
outperformed TPP for both read and write,
except for the slightly worse performance on platform C,
highlighting the advantage of asynchronous page migration in \textsc{Nomad}.

In the {\em stable} phase (i.e., {\em migration stable}), when most of the WSS data had been migrated from CXL memory or PM to the local DRAM, both \textsc{Nomad} and TPP achieved similar read/write bandwidth. This was because memory accesses were primarily served by the local DRAM with few page migrations, as shown in Table~\ref{tab:cam-ready-pagemigrationnumb}. Memtis performed the worst, achieving as low as 40\% of the performance of the other two approaches. We make two observations regarding Memtis's weaknesses. First, its stable phase performance is not drastically different from the transient phase. The migration statistics in Table~\ref{tab:cam-ready-pagemigrationnumb} show that Memtis performed significantly fewer page migrations. This explains its sub-optimal performance in the stable phase as most memory accesses were still served from slow memory. Second, a shorter cooling period in Memtis, which incentivizes more frequent migrations, led to better performance. This also suggests that sampling-based memory access tracking may not accurately identify and timely migrate hot pages to fast memory.

\smallskip

\noindent{\bf Medium WSS.} We increased the size of WSS and RSS to 13.5 GB and 27 GB, respectively. Similarly, we placed the first half of the RSS (13.5 GB) at the start of the local DRAM, followed by 2.5 GB of the WSS on the local DRAM, with the remaining 11 GB residing on CXL memory or PM. However, as the system (e.g., the OS kernel) required approximately 3-4 GB of memory, the WSS could barely fit in the fast tier, resulting in occasional and substantial migrations even during the stable phase. Accurately identifying hot pages and avoiding thrashing is crucial to achieving high performance for this medium-sized benchmark.

Unlike the small WSS case, Figures~\ref{fig:cam-ready-microbench-bw-platform-a} (b), \ref{fig:cam-ready-microbench-bw-platform-c} (b), and \ref{fig:cam-ready-microbench-bw-platform-d} (b) show that during the transient phase, \textsc{Nomad} and TPP generally achieved lower performance for both read and write compared to Memtis. This is because, under the medium WSS, the system experienced higher memory pressure than in the small WSS case, causing \textsc{Nomad} and TPP to conduct more page migrations (2x - 6x) and incur higher overhead than Memtis, as shown in Table ~\ref{tab:cam-ready-pagemigrationnumb}. Many of such migrations were futile during thrashing. Conversely, Memtis performed significantly fewer page migrations and avoided the waste. However, there was no evidence that Memtis effectively detected thrashing and throttled migration. The coarse-grained sampling was unable to accurately determine page temperature in a volatile situation and inadvertently sustained high performance under high memory pressure.

In the stable phase, \textsc{Nomad} significantly outperformed TPP in all cases, especially on platform D. These results show the benefit of \textsc{Nomad}'s transactional page migration and non-exclusive memory tiering compared to TPP's synchronous page migration and exclusive tiering. On platform D, which was equipped with an application-specific integrated circuit (ASIC)-based CXL memory implementation, the performance gap between fast and slow memory narrows. Thus, the software overhead associated with synchronous page migration was exacerbated and \textsc{Nomad} offered more pronounced performance gains. Similarly, \textsc{Nomad} achieved substantially higher performance in reads than Memtis and comparable performance in writes. Unlike in the small WSS case, in which asynchronous and transactional page migration in \textsc{Nomad} contributed most to its performance benefit, the advantage of page shadowing played a critical role in alleviating thrashing in the medium WSS case. Under memory thrashing, most demoted pages, which were recently promoted from the slow tier, can be simply discarded without migration. However, for write-intensive workloads, page shadowing requires one additional page fault for each write to restore a page's original read-write permission. This explains \textsc{Nomad}'s inferior write performance in the stable phase compared to Memtis.

\smallskip
\noindent{\bf Large WSS.} 
We scaled up the WSS and RSS both to 27 GB and fully populated local DRAM with the first 16 GB of the WSS. The remaining WSS spilled onto CXL memory or PM. Unlike the medium WSS that incurred intermittent memory thrashing, this workload caused continuous and severe thrashing as the size of hot data greatly exceeded the capacity of fast memory. 
Figures~\ref{fig:cam-ready-microbench-bw-platform-a} (c), \ref{fig:cam-ready-microbench-bw-platform-c} (c), and \ref{fig:cam-ready-microbench-bw-platform-d} (c) present the performance results in both the {\em transient} phase and the {\em stable} phase. Compared to the tests with the medium-sized workload in which \textsc{Nomad} could outperform Memtis for read-only benchmarks, especially in the stable phase, both \textsc{Nomad} and TPP performed worse than Memtis in almost all scenarios. It suggests that page fault-based tiered memory management, which makes per-page migration decisions upon access to a page, inevitably incurs high overhead during severe memory thrashing. 
Nevertheless, \textsc{Nomad} consistently and significantly outperformed TPP thanks to asynchronous, transactional page migration and page shadowing.

\begin{figure}[t]
    \centerline{\includegraphics[width=0.45\textwidth]{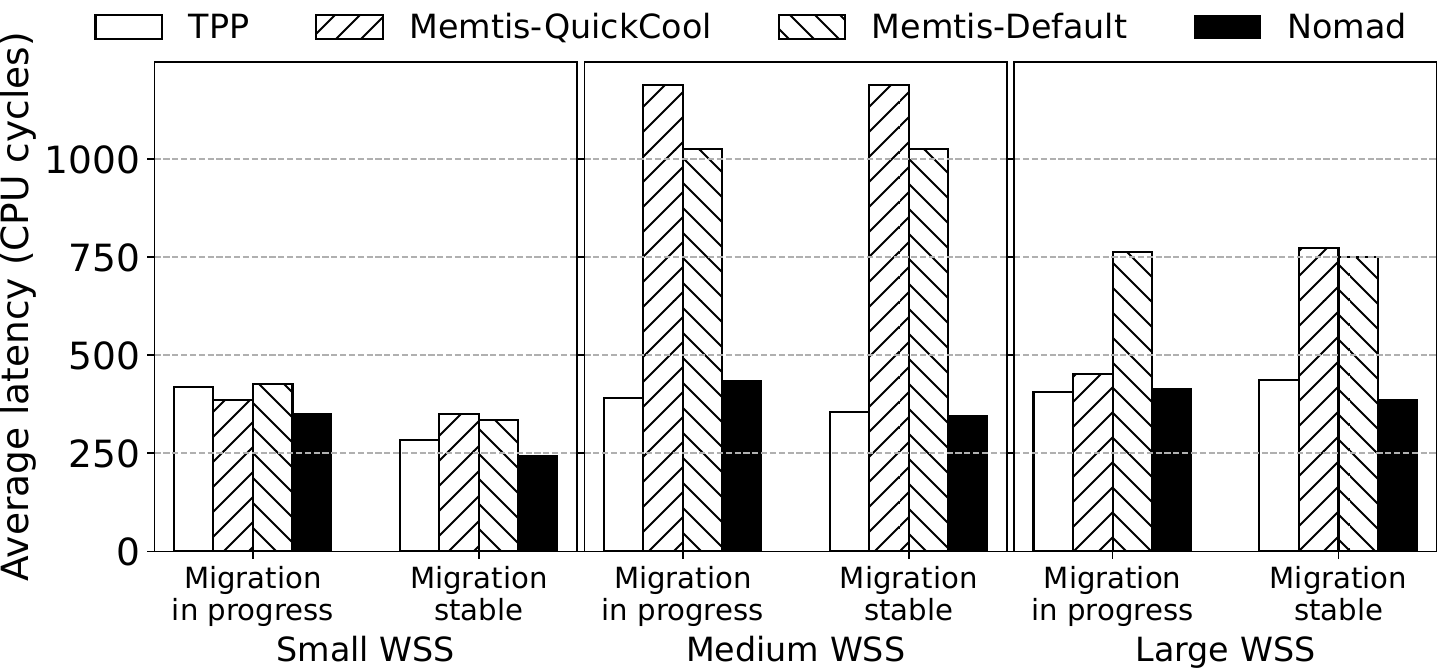}}
	\vspace{-1mm}
	\caption{The average cache line access latency on platform C. The benchmark is a point-chasing workload optimized for PEBS-based memory access tracking.  
     }
    \label{fig:cam-ready-microbench-llcmiss-lat}
	\vspace{-5mm}
\end{figure}

\smallskip
\noindent{\bf Limitations of PEBS-based approaches.} Our evaluation revealed several issues with PEBS-based memory access tracking. While Memtis prevented excessive migrations during thrashing, it achieved sub-optimal performance and failed to migrate all hot data to fast memory even when the WSS could fit in the fast tier. Due to the lack of hardware support for hot page tracking, PEBS-based approaches employ indirect metrics, such as LLC and TLB misses to sample recently accessed addresses to infer page temperature. Sampling-based memory tracking has two fundamental limitations. First, there is a difficult tradeoff between sampling rate and tracking accuracy. Second and most importantly, cache misses may not effectively capture hot pages. For most frequently accessed pages that always hit the caches, Memtis fails to collect enough (cache miss) samples to build the histogram. If such pages are evicted from the caches, e.g., due to conflict or coherence misses, they will be falsely regarded as ``cold'' pages. 

To demonstrate these limitations, we created a favorable scenario where Memtis can capture every page access. We used a pointer-chasing benchmark that repeatedly accesses multiple fixed-sized (1 GB) memory blocks. Within each 1 GB block, the benchmark randomly accesses all cache lines belonging to a block while accesses across blocks follow a Zipfian distribution. The number of blocks determines the WSS. Since the block size exceeds the LLC size in our testbeds, every access generates an LLC miss that can be captured by Memtis. Effective memory access tracking should identify hot blocks and place them in fast memory. 

Figure~\ref{fig:cam-ready-microbench-llcmiss-lat} shows the average latency to access a cache line in this benchmark on platform C. Note that platform C with PM was the only testbed on which Memtis has full tracking capability and can capture all the PEBS events. According to Table~\ref{tab:config}, a latency closer to DRAM performance ($\sim$250 cycles) indicates more effective page placement. As shown in Figure~\ref{fig:cam-ready-microbench-llcmiss-lat}, when the WSS exceeds fast tier capacity, Memtis achieved latency close to slow memory performance, suggesting that most hot pages still resided in the slow tier.  In comparison, page fault-based approaches, e.g., \textsc{Nomad} and TPP, can timely migrate hot pages and achieve low latency.

\begin{table}[t]
\centering
\footnotesize
  \label{tab:robustness}
  \begin{tabular}{c|c|c|c|c}
    \hline
    RSS & 23GB& 25GB& 27GB& 29GB \\
   \hline
   
    Total shadow page size & 3.93GB & 2.68GB & 2.2GB & 0.58GB \\
   \hline
\end{tabular}
  \caption{Shadow memory size as RSS changes on platform B. The size of tiered memory (DRAM+CXL) is 30.7 GB.}
  \vspace{-4mm}
  \label{robustness}
\end{table}

\smallskip

\noindent {\bf Robustness.} Page shadowing can potentially increase memory usage and in the worst case can cause OOM errors if shadow pages are not timely reclaimed. In this test, we evaluated \textsc{Nomad}'s shadow page reclamation. We measured the total memory usage and the size of shadow memory using a micro-benchmark that sequentially scans a predefined RSS area. Table~\ref{robustness} shows the change of shadow pages as we varied the RSS. The results suggest that \textsc{Nomad} effectively reclaimed shadow pages to reduce shadow memory usage as RSS increased and approached memory capacity.

\subsection{Real-world Applications}
We continued the evaluation of \textsc{Nomad} using three representative real-world applications with unique memory access patterns: Redis~\cite{redis}, PageRank~\cite{pagerankwiki}, and Liblinear~\cite{liblinear}. We ran these three applications on four platforms (as shown in Table~\ref{tab:config}) with two configurations: 1) a small RSS (under 32 GB) working with all platforms and 2) a large RSS (over 32 GB) only on platform C and D with large PM or CXL memory. In addition, we include results from a ``no migration'' baseline which disables page migrations to show whether tiered memory management is necessary. 

\begin{figure}[t]
	\centerline{\includegraphics[width=0.48\textwidth]{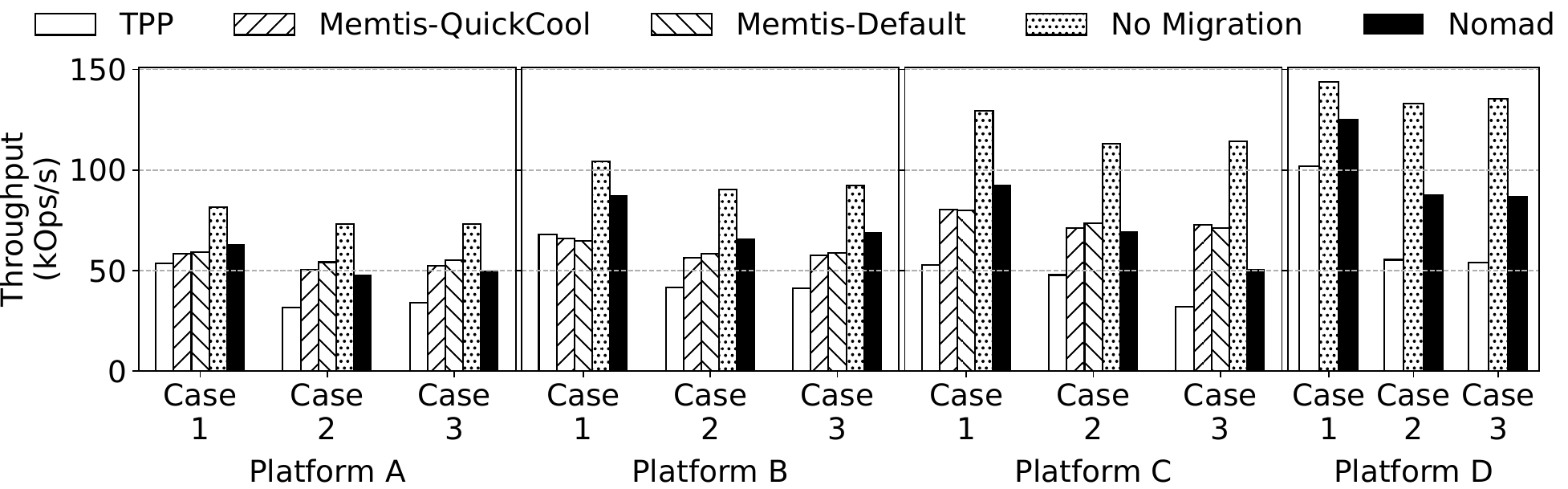}}
	\vspace{-1mm}
	\caption{Performance comparisons using Redis and YCSB between TPP, Memtis-Default, Memtis-QuickCool, \textsc{Nomad}, and ``no migration''. 
 }
    \label{fig:cam-ready-redis}
	\vspace{-4mm}
\end{figure}

\smallskip
\noindent{\bf Key-value store.} We first conducted experiments on a {\em latency-sensitive} key-value database, Redis~\cite{redis}. The workload was generated from YCSB~\cite{ycsb}, using its {\em update-heavy} workload A, with a 50/50 distribution of read and write operations. We crafted three cases with different RSS and total operations. Note that the parameters of YCSB were set as default unless otherwise specified. {Case 1}: We set {\em recordcount} to 6 million and {\em operationcount} to 8 million. After pre-loading the dataset, we used a customized tool to demote all memory pages to the slow tier before starting the experiment. The RSS of this case was 13GB. {Case 2}: We increased the RSS by setting {\em recordcount} to 10 million and {\em operationcount} to 12 million. We demoted all the memory pages to the slow tier in the same way. The RSS of this case was 24GB. {Case 3}: We kept the same total operations and RSS as Case 2. However, after pre-loading the dataset, we did {\em not} demote any memory pages.

Consistent with the micro-benchmarking results, Figure~\ref{fig:cam-ready-redis} shows that \textsc{Nomad} delivered superior performance (in terms of operations per second) compared to TPP across all platforms in all cases. In addition, \textsc{Nomad} outperformed Memtis when the WSS was small (i.e., in case 1), but suffered more performance degradation as the WSS increased (i.e., in case 2 and 3) due to an increased number of page migrations and additional overhead. Finally, all the page migration approaches underperformed compared to the ``no migration'' baseline. It is because the memory accesses generated by the YCSB workload were mostly ``random'', rendering migrating pages to the fast tier less effective, as those pages were unlikely to be accessed again. It indicates once again that page migration could incur nontrivial overhead, and a strategy to dynamically switch it on/off is needed.

\begin{figure*}[htbp]
    \centering
    \begin{minipage}[b]{0.3\textwidth}
        \centering
        \includegraphics[width=\textwidth]{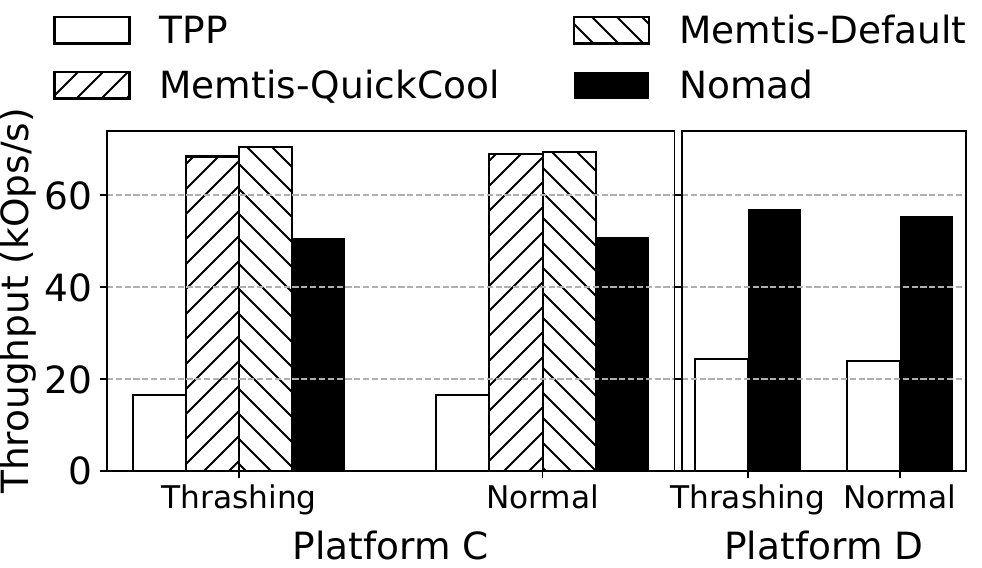}
        \vspace{-8mm}
        \caption{Redis (large RSS).} 
        \label{fig:cam-ready-huge-redis}
    \end{minipage}
    \hfill
    \begin{minipage}[b]{0.3\textwidth}
        \centering
        \includegraphics[width=\textwidth]{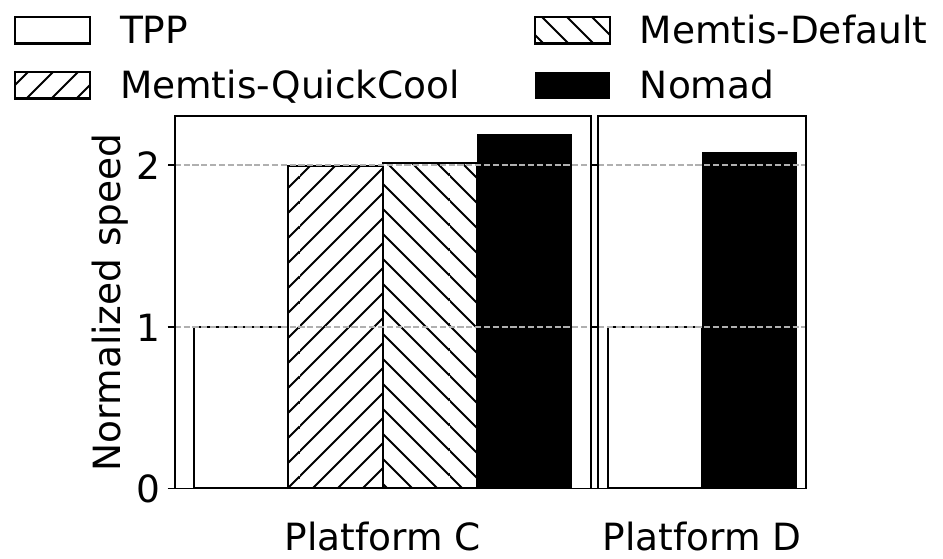}
        \vspace{-8mm}
        \caption{Page ranking (large RSS). }
        \label{fig:cam-ready-huge-pageranking}
    \end{minipage}
    \hfill
    \begin{minipage}[b]{0.3\textwidth}
        \centering
        \includegraphics[width=\textwidth]{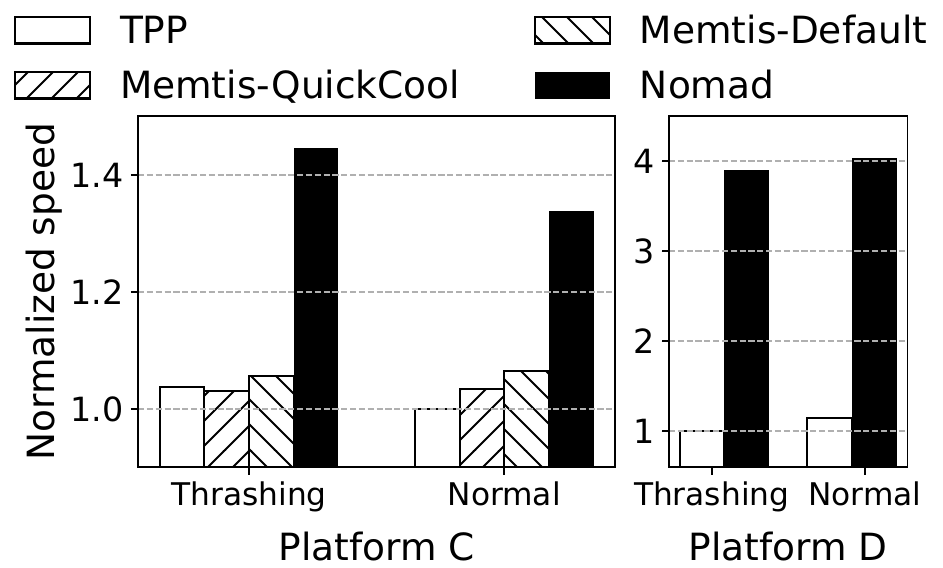}
        \vspace{-8mm}
        \caption{Liblinear (large RSS).} 
        \label{fig:cam-ready-huge-liblinear}
    \end{minipage}
    \vspace{-4mm}
\end{figure*}

We further increased the RSS of the database and operations of YCSB by setting the {\em recordcount} to 20 million and {\em operationcount} to 30 million. The RSS for this case was 36.5GB, exceeding the total size of the tiered memory on platforms A and B. Thus, the large RSS test was only performed on platforms C and D. We tested two initial memory placement strategies for the database -- 1) {\em thrashing} that allocated all pages first to the slow tier and immediately invoked intensive page migrations, and 2) {\em normal} that prioritized page allocation to fast memory and triggered page migration only under memory pressure. As shown in Figure~\ref{fig:cam-ready-huge-redis}, \textsc{Nomad} outperformed TPP due to its graceful performance degradation during thrashing but fell short of matching Memtis's performance. The initial placement strategy did not substantially affect the results and performance under different placements eventually converged.

\begin{figure}[t]
\centerline{\includegraphics[width=0.45\textwidth]{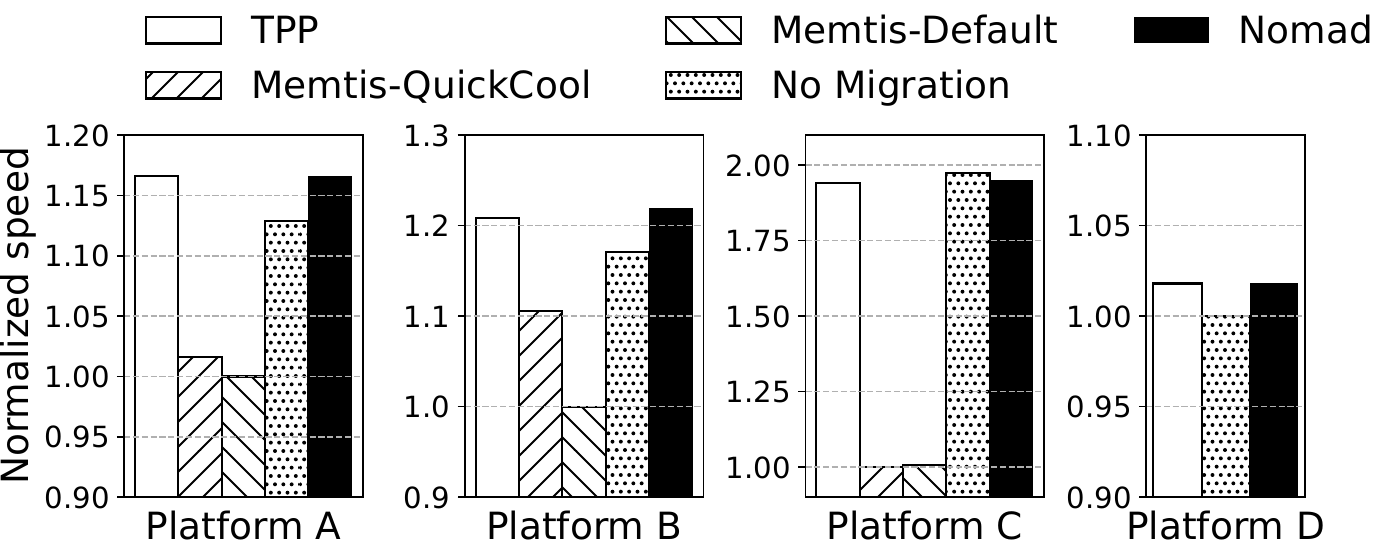}}
	\vspace{-1mm}
	\caption{Performance comparisons of PageRank between non-migration, TPP, Memtis, and \textsc{Nomad}. Performance is normalized to the approach with the lowest speed.}
	\vspace{-4mm}
    \label{fig:cam-ready-pageranking}
\end{figure}

\smallskip
\noindent{\bf Graph-based computation.} We used PageRank~\cite{pagerankwiki}, an application used to rank web pages. It involves iterative computations to determine the rank of each page, based on the link structures of the entire web. As the size of the dataset increases, the computational complexity also increases, making it both memory-intensive and compute-intensive. We used a benchmark suite~\cite{pagerank} to generate a synthetic uniform-random graph comprising $2^{26}$ vertices, each with an average of 20 edges. The RSS in this experiment was 22 GB, indicating that the memory pages were distributed at both the local DRAM and remote CXL memory or PM.

Figure~\ref{fig:cam-ready-pageranking} illustrates that there was negligible variance in performance between scenarios with page migrations (using \textsc{Nomad} and TPP) and without page migrations (no migration). The results suggest that: 1) For {\em non-latency-sensitive} applications, such as PageRank, using CXL memory can significantly expand the local DRAM capacity without adversely impacting application-level performance. 2) In such scenarios, page migration appears to be unnecessary. 
These findings also reveal that the overhead associated with \textsc{Nomad}'s page migration minimally influences PageRank's performance. Additionally, it was observed that among all evaluated scenarios, Memtis exhibited the least efficient performance.

Figure~\ref{fig:cam-ready-huge-pageranking} shows the case when we scaled the RSS to a very large scale on platforms C \& D. When the PageRank program started, it first used up to 100GB memory, then its RSS size dropped to 45GB to 50GB. \textsc{Nomad} achieved 2x the performance of TPP (both platforms) and slightly better than Memtis (platform C), due to more frequent page migrations -- the local DRAM (16 GB) was not large enough to accommodate the WSS in this case. 


\begin{figure}[t]
\centerline{\includegraphics[width=0.42\textwidth]{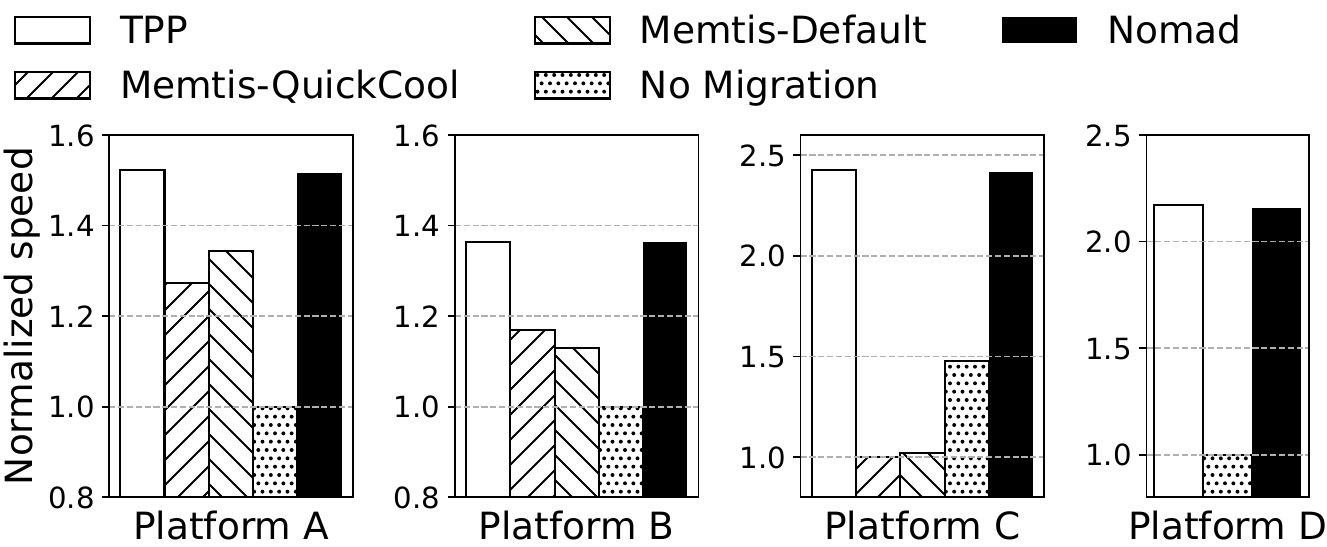}}
	\vspace{-1mm}
	\caption{Performance comparisons of Liblinear between non-migration, TPP, Memtis, and \textsc{Nomad}. Performance is normalized to the approach with the lowest speed.}
	\vspace{-1mm}
 \label{fig:cam-ready-liblinear}
\end{figure}

\smallskip
\noindent{\bf Machine learning.} Our final evaluation of \textsc{Nomad} involved using the machine learning library Liblinear~\cite{liblinear}, known for its large-scale linear classification capabilities. We executed Liblinear with an L1 regularized logistic regression workload
with an RSS of 10 GB. Prior to each execution, we used our tool to demote all memory pages associated with the Liblinear workload to the slower memory tier.

Figure~\ref{fig:cam-ready-liblinear} demonstrates that both \textsc{Nomad} and TPP significantly outperformed ``no migration'' and Memtis across all platforms, with performance improvement ranging from 20\% to 150\%. This result further illustrates that when the WSS is smaller than the local DRAM, \textsc{Nomad} and TPP can substantially enhance application performance by timely migrating hot pages to the faster memory tier. Figure~\ref{fig:cam-ready-huge-liblinear} shows that with a much larger model and RSS when running Liblinear, NOMAD consistently achieved high performance across all cases. In contrast, TPP's performance significantly declined, likely due to inefficiency issues, as frequent, high bursts in kernel CPU time were observed during TPP execution.

 

\begin{table}[t]
\centering
\small
  
  \begin{tabular}{c||c}
    \hline
    Workload type & Success : Aborted \\
   \hline
   \hline
     Liblinear (large RSS) on platform C &  1:1.9 \\
   \hline
     Liblinear (large RSS) on platform D & 2.6:1  \\
   \hline
    Redis (large RSS) on platform C & 153:1  \\
   \hline
    Redis (large RSS) on platform D &  278.2:1  \\
   \hline
\end{tabular}
  \caption{The Success rate of transactional migration.}
  \vspace{-4mm}
     \label{tab:txAppSuccessRateCompare}
\end{table}

\smallskip
\noindent{\bf Migration success rate.} 
As stated in Section~\ref{tpm}, \textsc{Nomad}'s transactional page migration may be aborted due to updates to the migrating page, resulting in wasted memory bandwidth and CPU cycles. Subsequent retries could also fail. A low success rate could negatively affect application performance. Table~\ref{tab:txAppSuccessRateCompare} shows \textsc{Nomad}'s migration success rate for Liblinear and Redis on platforms C and D. We chose a large RSS for both applications and ensured there were sufficient cross-tier migrations.  
We observed a low success rate for Liblinear while Redis had a high success rate. Interestingly, this contrasted with \textsc{Nomad}’s performance -- it was excellent with Liblinear but poor with Redis with large RSS. This suggests that a high success rate in page migrations does not necessarily lead to high performance. A low success rate indicates that the pages being migrated by \textsc{Nomad} are also being modified by other processes, implying their ``hotness''. Timely migration of such pages can benefit ongoing and future accesses. 

\noindent{\bf Summary.}
The results from micro-benchmarks and applications indicate that when the WSS was smaller than the performance tier, \textsc{Nomad} enabled workloads to maintain higher performance than Memtis through asynchronous, transactional page migrations. However, when the WSS was comparable to or exceeded the performance tier capacity, leading to memory thrashing, the page-fault-based migration in \textsc{Nomad} became detrimental to workload performance, underperforming Memtis in write operations. Notably, \textsc{Nomad}'s page shadowing feature preserved the efficiency of read operations even under severe memory thrashing, often maintaining comparable or superior performance to Memtis. In all test scenarios, \textsc{Nomad} significantly outperformed the state-of-the-art page-fault-based migration approach, TPP.

The evaluation results across four different platforms reveal the following observations:
First, \textsc{Nomad} generally performed better on platform D, which was equipped with faster and larger CXL memory, compared to other platforms. Additionally, the reduced performance gap between fast and slow memory on platform D allowed \textsc{Nomad} to achieve greater performance gains than TPP, as the performance overhead from TPP's synchronous page migration was more pronounced.
Second, while platforms A and B showed similar behavior in micro-benchmarks, their application-level performance varied (slightly) across different applications, suggesting that specific CPU features (differing between the off-the-shelf Intel Sapphire Rapids CPU for platform A and the engineering sample for platform B) may affect the performance of page migration under more realistic workloads. 

\section{Discussions and Future Work}
\label{sec:discussion}
The key insight from \textsc{Nomad}'s evaluation is that page migration, especially under memory pressure, has a detrimental impact on overall application performance. While \textsc{Nomad} achieved graceful performance degradation and much higher performance than TPP, an approach based on synchronous page migration, its performance is sub-optimal compared to that without page migration. When the program's working set exceeds the capacity of the fast tier, the most effective strategy is to access pages directly from their initial placement, completely disabling page migration. It is straightforward to detect memory thrashing, e.g., frequent and equal number of page demotions and promotions, and disable page migrations. However, estimating the working set size to resume page migration becomes challenging, as the working set now spans multiple tiers. It requires global memory tracking, which could be prohibitively expensive, to identify the hot data set that can potentially be migrated to the fast tier. We plan to extend \textsc{Nomad} to unilaterally throttle page promotions and monitor page demotions to effectively manage memory pressure on the fast tier. Note that this would require the development of a new page migration policy, which is orthogonal to the \textsc{Nomad} page migration mechanisms proposed in this work.

Impact of Platform Characteristics: There exist difficult tradeoffs between page fault-based access tracking, such as TPP and \textsc{Nomad}, and hardware performance counter-based memory access sampling like Memtis. While page fault-based tracking effectively captures access recency, it can be potentially expensive and on the critical path of program execution. In comparison, hardware-based access sampling is off the critical path and captures access frequency. However, it is not responsive to workload changes and its accuracy relies on the sampling rate. One advantage of \textsc{Nomad} is that it is a page fault-based migration approach that is asynchronous and off the critical path. A potential future work is integrating \textsc{Nomad} with hardware-based, access frequency tracking, such as Memtis, to enhance the current migration policy.

\section{Conclusion}
\label{sec:conclusion}

This paper introduces non-exclusive memory tiering as an alternative to the common exclusive memory tiering strategy, where each page is confined to either fast or slow memory. The proposed approach, implemented in \textsc{Nomad}, leverages transactional page migration and page shadowing to enhance page management in Linux. Unlike traditional page migration, \textsc{Nomad} ensures asynchronous migration and retains shadow copies of recently promoted pages. Through comprehensive evaluations, \textsc{Nomad} demonstrates up to 6x performance improvement over existing methods, addressing critical performance degradation issues in exclusive memory tiering, especially under memory pressure. The paper calls for further research in tiered memory-aware memory allocation.

\section{Acknowledgments}
We thank our shepherd, Sudarsun Kannan, and the anonymous reviewers for their constructive feedback. This work was supported in part by NSF grants CCF-1845706, CNS-2415774, CCF-2415473, and the gift and equipment from Intel Labs and Micron Technology.

\newpage

%% file: osdi24_ae_appendix_template.tex


\appendix
\section{Artifact Appendix}

\subsection*{Abstract}
The artifact contains the source code of \textsc{Nomad}, TPP, and Memtis for
reproducing the results and graphs presented in the paper. The code works on platforms with Persistent Memory, Intel Agilex CXL memory, and/or Micron CXL memory. To facilitate the reproduction, we have provided a collection of scripts for compiling and installing these approaches, executing the experiments, collecting logs, and creating graphs. More details are available in the "README.md" file.
\subsection*{Scope}


This artifact demonstrates \textsc{Nomad}'s strengths and weaknesses over TPP and Memtis across various scenarios and platforms, as elaborated in the Evaluation section. 

It is open-source and can be used for further research, development, or other purposes by the community.

\subsection*{Contents}
\textbf{\textsc{Nomad}, TPP and Memtis implementation.} 
We provide two separate patches to enable \textsc{Nomad} and TPP to work with the upstream kernel version v5.13-rc6. In particular, the TPP patch comes from the Linux community email discussions. Memtis, on the other hand, is directly incorporated from its original artifact, with a few minor bugs fixed.
\smallskip 

\noindent\textbf{Documentation} 
The "Reproducing Paper Results" section of "README.md" provides a step-by-step guide for reproducing the results in the paper. This guide includes instructions for compiling the three implementations (i.e., \textsc{Nomad}, TPP, and Memtis), running the experiments, and generating the graphs as presented in the paper.

\subsection*{Hosting}


Artifact link: \href{https://github.com/lingfenghsiang/Nomad}{https://github.com/lingfenghsiang/Nomad} \\
Artifact license: GNU GPL V3.0 \\
Artifact version tag: v0.0

\subsection*{Requirements}
To reproduce the results in the paper, the system under test requires one NUMA node with a CPU and another CPU-less NUMA node. If the system has more NUMA nodes, the operating system might encounter unexpected errors. Additionally, Memtis is only fully functional on platforms with Optane Persistent Memory. More details are included in the "Prerequisites" section of "README.md".





%% file: usenix.bbl
\begin{thebibliography}{10}

\bibitem{cxl}
\url{https://www.computeexpresslink.org/}.

\bibitem{autoNUMA}
Autonuma: the other approach to numa scheduling.
\newblock \url{https://lwn.net/Articles/488709/}.

\bibitem{damon}
Damon-based reclamation.
\newblock \url{https://docs.kernel.org/admin-guide/mm/damon/reclaim.html#:~:text=DAMON%2Dbased%20Reclamation%20(DAMON_RECLAIM),of%20memory%20pressure%20and%20requirements.}

\bibitem{hbm}
https://blocksandfiles.com/2023/11/20/accelerating-high-bandwidth-memory-to-light-speed/.
\newblock \url{https://blocksandfiles.com/2023/11/20/accelerating-high-bandwidth-memory-to-light-speed/}.

\bibitem{liblinear}
https://www.csie.ntu.edu.tw/~cjlin/libsvmtools/multicore-liblinear/.

\bibitem{agilex}
Intel agilex® 7 fpga and soc fpga.
\newblock \url{https://www.intel.com/content/www/us/en/products/details/fpga/agilex/7.html}.

\bibitem{optane}
Intel optane dimm.
\newblock \url{https://www.intel.com/content/www/us/en/architecture-and-technology/optane-dc-persistent-memory.html}.

\bibitem{pagerank}
Pagerank.
\newblock \url{https://github.com/sbeamer/gapbs}.

\bibitem{pagerankwiki}
Pagerank wiki.
\newblock \url{https://en.wikipedia.org/wiki/PageRank}.

\bibitem{redis}
Redis.
\newblock \url{https://redis.io/}.

\bibitem{ycsb}
Ycsb.
\newblock \url{https://github.com/brianfrankcooper/YCSB}.

\bibitem{266253}
Caching less for better performance: Balancing cache size and update cost of flash memory cache in hybrid storage systems.
\newblock In {\em 10th USENIX Conference on File and Storage Technologies (FAST 12)\/} (San Jose, CA, Feb. 2012), USENIX Association.

\bibitem{abulila2019flatflash}
{\sc Abulila, A., Mailthody, V.~S., Qureshi, Z., Huang, J., Kim, N.~S., Xiong, J., and Hwu, W.-m.}
\newblock Flatflash: Exploiting the byte-accessibility of ssds within a unified memory-storage hierarchy.
\newblock In {\em Proceedings of the Twenty-Fourth International Conference on Architectural Support for Programming Languages and Operating Systems\/} (2019), pp.~971--985.

\bibitem{thermostat}
{\sc Agarwal, N., and Wenisch, T.~F.}
\newblock Thermostat: Application-transparent page management for two-tiered main memory.
\newblock {\em SIGPLAN Not. 52}, 4 (apr 2017), 631–644.

\bibitem{8714805}
{\sc Ahmadian, S., Salkhordeh, R., and Asadi, H.}
\newblock Lbica: A load balancer for i/o cache architectures.
\newblock In {\em 2019 Design, Automation \& Test in Europe Conference \& Exhibition (DATE)\/} (2019), pp.~1196--1201.

\bibitem{Alian-micro21}
{\sc Alian, M., Shin, J., Kang, K.-D., Wang, R., Daglis, A., Kim, D., and Kim, N.~S.}
\newblock Idio: Orchestrating inbound network data on server processors.
\newblock {\em IEEE Computer Architecture Letters 20}, 1 (2021), 30--33.

\bibitem{amd_tlb_broadcast}
{\sc AMD}.
\newblock Zynq ultrascale+ device technical reference manual (ug1085), 2023.
\newblock \url{https://docs.amd.com/r/en-US/ug1085-zynq-ultrascale-trm/TLB-Maintenance-Operations}.

\bibitem{arm_tlb_broadcast}
{\sc ARM}.
\newblock Learn the architecture - aarch64 memory management guide, 2024.
\newblock \url{https://developer.arm.com/documentation/101811/0103/Translation-Lookaside-Buffer-maintenance/Format-of-a-TLB-operation}.

\bibitem{Bergman_22ismm}
{\sc Bergman, S., Faldu, P., Grot, B., Vilanova, L., and Silberstein, M.}
\newblock Reconsidering os memory optimizations in the presence of disaggregated memory.
\newblock In {\em Proceedings of the 2022 ACM SIGPLAN International Symposium on Memory Management\/} (New York, NY, USA, 2022), ISMM 2022, Association for Computing Machinery, p.~1–14.

\bibitem{concurrent}
{\sc Bock, S., Childers, B.~R., Melhem, R., and Moss\'{e}, D.}
\newblock Concurrent page migration for mobile systems with os-managed hybrid memory.
\newblock In {\em Proceedings of the 11th ACM Conference on Computing Frontiers\/} (New York, NY, USA, 2014), CF '14, Association for Computing Machinery.

\bibitem{270689}
{\sc Burnett, N.~C., Bent, J., Arpaci-Dusseau, A.~C., and Arpaci-Dusseau, R.~H.}
\newblock Exploiting {Gray-Box} knowledge of {Buffer-Cache} management.
\newblock In {\em 2002 USENIX Annual Technical Conference (USENIX ATC 02)\/} (Monterey, CA, June 2002), USENIX Association.

\bibitem{chen2011hystor}
{\sc Chen, F., Koufaty, D.~A., and Zhang, X.}
\newblock Hystor: Making the best use of solid state drives in high performance storage systems.
\newblock In {\em Proceedings of the international conference on Supercomputing\/} (2011), pp.~22--32.

\bibitem{batman}
{\sc Chou, C., Jaleel, A., and Qureshi, M.}
\newblock Batman: Techniques for maximizing system bandwidth of memory systems with stacked-dram.
\newblock In {\em Proceedings of the International Symposium on Memory Systems\/} (New York, NY, USA, 2017), MEMSYS '17, Association for Computing Machinery, p.~268–280.

\bibitem{Padmapriya_23asplos}
{\sc Duraisamy, P., Xu, W., Hare, S., Rajwar, R., Culler, D., Xu, Z., Fan, J., Kennelly, C., McCloskey, B., Mijailovic, D., Morris, B., Mukherjee, C., Ren, J., Thelen, G., Turner, P., Villavieja, C., Ranganathan, P., and Vahdat, A.}
\newblock Towards an adaptable systems architecture for memory tiering at warehouse-scale.
\newblock In {\em Proceedings of the 28th ACM International Conference on Architectural Support for Programming Languages and Operating Systems, Volume 3\/} (New York, NY, USA, 2023), ASPLOS 2023, Association for Computing Machinery, p.~727–741.

\bibitem{270754}
{\sc Forney, B.~C., and Arpaci-Dusseau, A.~C.}
\newblock {Storage-Aware} caching: Revisiting caching for heterogeneous storage systems.
\newblock In {\em Conference on File and Storage Technologies (FAST 02)\/} (Monterey, CA, Jan. 2002), USENIX Association.

\bibitem{267006}
{\sc Guerra, J., Pucha, H., Glider, J., Belluomini, W., and Rangaswami, R.}
\newblock Cost effective storage using extent based dynamic tiering.
\newblock In {\em 9th USENIX Conference on File and Storage Technologies (FAST 11)\/} (San Jose, CA, Feb. 2011), USENIX Association.

\bibitem{180194}
{\sc Holland, D.~A., Angelino, E., Wald, G., and Seltzer, M.~I.}
\newblock Flash caching on the storage client.
\newblock In {\em 2013 USENIX Annual Technical Conference (USENIX ATC 13)\/} (San Jose, CA, June 2013), USENIX Association, pp.~127--138.

\bibitem{intelcpurpc}
{\sc Intel{\textregistered}}.
\newblock Remote action request --white paper.
\newblock {\em revision 1.0\/} (2021).
\newblock \url{https://www.intel.com/content/dam/develop/external/us/en/documents/341431-remote-action-request-white-paper.pdf}.

\bibitem{269067}
{\sc Jiang, S., Ding, X., Chen, F., Tan, E., and Zhang, X.}
\newblock {DULO}: An effective buffer cache management scheme to exploit both temporal and spatial localities.
\newblock In {\em 4th USENIX Conference on File and Storage Technologies (FAST 05)\/} (San Francisco, CA, Dec. 2005), USENIX Association.

\bibitem{jun2017hbm}
{\sc Jun, H., Cho, J., Lee, K., Son, H.-Y., Kim, K., Jin, H., and Kim, K.}
\newblock Hbm (high bandwidth memory) dram technology and architecture.
\newblock In {\em 2017 IEEE International Memory Workshop (IMW)\/} (2017), IEEE, pp.~1--4.

\bibitem{pcieovercxl}
{\sc Jung, M.}
\newblock Hello bytes, bye blocks: Pcie storage meets compute express link for memory expansion (cxl-ssd).
\newblock HotStorage '22, Association for Computing Machinery, p.~45–51.

\bibitem{Jonghyeon_21fast}
{\sc Kim, J., Choe, W., and Ahn, J.}
\newblock Exploring the design space of page management for {Multi-Tiered} memory systems.
\newblock In {\em 2021 USENIX Annual Technical Conference (USENIX ATC 21)\/} (July 2021), USENIX Association, pp.~715--728.

\bibitem{6005391}
{\sc Kim, Y., Gupta, A., Urgaonkar, B., Berman, P., and Sivasubramaniam, A.}
\newblock Hybridstore: A cost-efficient, high-performance storage system combining ssds and hdds.
\newblock In {\em 2011 IEEE 19th Annual International Symposium on Modelling, Analysis, and Simulation of Computer and Telecommunication Systems\/} (2011), pp.~227--236.

\bibitem{180727}
{\sc Koller, R., Marmol, L., Rangaswami, R., Sundararaman, S., Talagala, N., and Zhao, M.}
\newblock Write policies for host-side flash caches.
\newblock In {\em 11th USENIX Conference on File and Storage Technologies (FAST 13)\/} (San Jose, CA, Feb. 2013), USENIX Association, pp.~45--58.

\bibitem{7266934}
{\sc Koller, R., Mashtizadeh, A.~J., and Rangaswami, R.}
\newblock Centaur: Host-side ssd caching for storage performance control.
\newblock In {\em 2015 IEEE International Conference on Autonomic Computing\/} (2015), pp.~51--60.

\bibitem{kwon2017strata}
{\sc Kwon, Y., Fingler, H., Hunt, T., Peter, S., Witchel, E., and Anderson, T.}
\newblock Strata: A cross media file system.
\newblock In {\em Proceedings of the 26th Symposium on Operating Systems Principles\/} (2017), pp.~460--477.

\bibitem{memtis}
{\sc Lee, T., Monga, S.~K., Min, C., and Eom, Y.~I.}
\newblock Memtis: Efficient memory tiering with dynamic page classification and page size determination.
\newblock In {\em Proceedings of the 29th Symposium on Operating Systems Principles\/} (New York, NY, USA, 2023), SOSP '23, Association for Computing Machinery, p.~17–34.

\bibitem{li2023pond}
{\sc Li, H., Berger, D.~S., Hsu, L., Ernst, D., Zardoshti, P., Novakovic, S., Shah, M., Rajadnya, S., Lee, S., Agarwal, I., et~al.}
\newblock Pond: Cxl-based memory pooling systems for cloud platforms.
\newblock In {\em Proceedings of the 28th ACM International Conference on Architectural Support for Programming Languages and Operating Systems, Volume 2\/} (2023), pp.~574--587.

\bibitem{p2cache}
{\sc Lin, Z., Xiang, L., Rao, J., and Lu, H.}
\newblock {P2CACHE}: Exploring tiered memory for {In-Kernel} file systems caching.
\newblock In {\em 2023 USENIX Annual Technical Conference (USENIX ATC 23)\/} (Boston, MA, July 2023), USENIX Association, pp.~801--815.

\bibitem{288741}
{\sc Lin, Z., Xiang, L., Rao, J., and Lu, H.}
\newblock {P2CACHE}: Exploring tiered memory for {In-Kernel} file systems caching.
\newblock In {\em 2023 USENIX Annual Technical Conference (USENIX ATC 23)\/} (Boston, MA, July 2023), USENIX Association, pp.~801--815.

\bibitem{6557143}
{\sc Liu, K., Zhang, X., Davis, K., and Jiang, S.}
\newblock Synergistic coupling of ssd and hard disk for qos-aware virtual memory.
\newblock In {\em 2013 IEEE International Symposium on Performance Analysis of Systems and Software (ISPASS)\/} (2013), pp.~24--33.

\bibitem{Loh2012ChallengesIH}
{\sc Loh, G.~H., Jayasena, N., Chung, J., Reinhardt, S.~K., O'Connor, J.~M., and McGrath, K.~J.}
\newblock Challenges in heterogeneous die-stacked and off-chip memory systems.

\bibitem{Adnan_22hpca}
{\sc Maruf, A., Ghosh, A., Bhimani, J., Campello, D., Rudoff, A., and Rangaswami, R.}
\newblock Multi-clock: Dynamic tiering for hybrid memory systems.
\newblock In {\em 2022 IEEE International Symposium on High-Performance Computer Architecture (HPCA)\/} (Los Alamitos, CA, USA, apr 2022), IEEE Computer Society, pp.~925--937.

\bibitem{Maruf_23asplos_tpp}
{\sc Maruf, H.~A., Wang, H., Dhanotia, A., Weiner, J., Agarwal, N., Bhattacharya, P., Petersen, C., Chowdhury, M., Kanaujia, S., and Chauhan, P.}
\newblock Tpp: Transparent page placement for cxl-enabled tiered-memory.
\newblock In {\em Proceedings of the 28th ACM International Conference on Architectural Support for Programming Languages and Operating Systems, Volume 3\/} (New York, NY, USA, 2023), ASPLOS 2023, Association for Computing Machinery, p.~742–755.

\bibitem{7056027}
{\sc Meswani, M.~R., Blagodurov, S., Roberts, D., Slice, J., Ignatowski, M., and Loh, G.~H.}
\newblock Heterogeneous memory architectures: A hw/sw approach for mixing die-stacked and off-package memories.
\newblock In {\em 2015 IEEE 21st International Symposium on High Performance Computer Architecture (HPCA)\/} (2015), pp.~126--136.

\bibitem{papagiannis2020optimizing}
{\sc Papagiannis, A., Xanthakis, G., Saloustros, G., Marazakis, M., and Bilas, A.}
\newblock Optimizing memory-mapped $\{$I/O$\}$ for fast storage devices.
\newblock In {\em 2020 USENIX Annual Technical Conference (USENIX ATC 20)\/} (2020), pp.~813--827.

\bibitem{ruan2020aifm}
{\sc Ruan, Z., Schwarzkopf, M., Aguilera, M.~K., and Belay, A.}
\newblock $\{$AIFM$\}$:$\{$High-Performance$\}$,$\{$Application-Integrated$\}$ far memory.
\newblock In {\em 14th USENIX Symposium on Operating Systems Design and Implementation (OSDI 20)\/} (2020), pp.~315--332.

\bibitem{6493624}
{\sc Sim, J., Loh, G.~H., Kim, H., OConnor, M., and Thottethodi, M.}
\newblock A mostly-clean dram cache for effective hit speculation and self-balancing dispatch.
\newblock In {\em 2012 45th Annual IEEE/ACM International Symposium on Microarchitecture\/} (2012), pp.~247--257.

\bibitem{10.5555/1855511.1855519}
{\sc Soundararajan, G., Prabhakaran, V., Balakrishnan, M., and Wobber, T.}
\newblock Extending ssd lifetimes with disk-based write caches.
\newblock In {\em Proceedings of the 8th USENIX Conference on File and Storage Technologies\/} (USA, 2010), FAST'10, USENIX Association, p.~8.

\bibitem{Sun2023DemystifyingCM}
{\sc Sun, Y., Yuan, Y., Yu, Z., Kuper, R., Jeong, I., Wang, R., and Kim, N.~S.}
\newblock Demystifying cxl memory with genuine cxl-ready systems and devices.
\newblock {\em ArXiv abs/2303.15375\/} (2023).

\bibitem{wu2021storage}
{\sc Wu, K., Guo, Z., Hu, G., Tu, K., Alagappan, R., Sen, R., Park, K., Arpaci-Dusseau, A.~C., and Arpaci-Dusseau, R.~H.}
\newblock The storage hierarchy is not a hierarchy: Optimizing caching on modern storage devices with orthus.
\newblock In {\em 19th USENIX Conference on File and Storage Technologies (FAST 21)\/} (2021), pp.~307--323.

\bibitem{5581609}
{\sc Wu, X., and Reddy, A.~N.}
\newblock Exploiting concurrency to improve latency and throughput in a hybrid storage system.
\newblock In {\em 2010 IEEE International Symposium on Modeling, Analysis and Simulation of Computer and Telecommunication Systems\/} (2010), pp.~14--23.

\bibitem{xiang2022characterizing}
{\sc Xiang, L., Zhao, X., Rao, J., Jiang, S., and Jiang, H.}
\newblock Characterizing the performance of intel optane persistent memory: a close look at its on-dimm buffering.
\newblock In {\em Proceedings of the Seventeenth European Conference on Computer Systems\/} (2022), pp.~488--505.

\bibitem{Yan_19asplos_nimble}
{\sc Yan, Z., Lustig, D., Nellans, D., and Bhattacharjee, A.}
\newblock Nimble page management for tiered memory systems.
\newblock In {\em Proceedings of the Twenty-Fourth International Conference on Architectural Support for Programming Languages and Operating Systems\/} (New York, NY, USA, 2019), ASPLOS '19, Association for Computing Machinery, p.~331–345.

\bibitem{optane_study}
{\sc Yang, J., Kim, J., Hoseinzadeh, M., Izraelevitz, J., and Swanson, S.}
\newblock An empirical guide to the behavior and use of scalable persistent memory.
\newblock In {\em Proceedings of the 18th USENIX Conference on File and Storage Technologies\/} (USA, 2020), FAST'20, USENIX Association, p.~169–182.

\bibitem{pmperf}
{\sc Yang, J., Kim, J., Hoseinzadeh, M., Izraelevitz, J., and Swanson, S.}
\newblock An empirical guide to the behavior and use of scalable persistent memory.
\newblock In {\em 18th USENIX Conference on File and Storage Technologies (FAST 20)\/} (Santa Clara, CA, Feb. 2020), USENIX Association, pp.~169--182.

\bibitem{5749736}
{\sc Yang, Q., and Ren, J.}
\newblock I-cash: Intelligently coupled array of ssd and hdd.
\newblock In {\em 2011 IEEE 17th International Symposium on High Performance Computer Architecture\/} (2011), pp.~278--289.

\end{thebibliography}
